\documentclass[twocol]{ametsocV6.1}

\usepackage[english]{babel}
\usepackage[T1]{fontenc}
\usepackage[utf8]{inputenc}
\usepackage{csquotes}
\usepackage{amsmath,amssymb}
\usepackage{graphicx}
\usepackage{float}
\usepackage{subfig}
\usepackage{subfloat}
\usepackage{verbatim}

\usepackage{acro}
\DeclareAcronym{enso}{
  short=ENSO,
  long=El Niño Southern Oscillation,
  long-format=\itshape,
}
\DeclareAcronym{nao}{
  short=NAO,
  long=North Atlantic Oscillation,
  long-format=\itshape,
}
\DeclareAcronym{ivp}{
  short=IVP,
  long=initial value problem,
}
\DeclareAcronym{cdf}{
  short=CDF,
  long=cumulative distribution function,
}

\title{Committor Functions for Climate Phenomena at the Predictability Margin: The Example of El Niño Southern Oscillation in the Jin and Timmermann Model}
\authors{Dario Lucente\aff{a}, Corentin Herbert\aff{a}, Freddy Bouchet\aff{a}\correspondingauthor{Freddy Bouchet, Freddy.Bouchet@ens-lyon.fr}}

\affiliation{\aff{a}{ENSL, CNRS, Laboratoire de Physique, F-69342 Lyon, France}}

\abstract{
  Many atmosphere and climate phenomena lie in the gray zone between weather and climate: they are not amenable to deterministic forecast, but they still depend on the initial condition.
  A natural example is medium-range forecasting, which is inherently probabilistic because it lies beyond the deterministic predictability time of the atmosphere, but for which statistically significant prediction can be made which depend on the current state of the system.
  Similarly, one may ask the probability of occurrence of an El Niño event several months ahead of time.
We introduce a quantity which corresponds precisely to this type of prediction problem: the committor function is the probability that an event takes place within a given time window, as a function of the initial condition.
 We compute it in the case of a low-dimensional stochastic model for El-Ni\~no, the Jin and Timmermann model.
  In this context, we show that the ability to predict the probability of occurrence of the event of interest may differ strongly depending on the initial state.
  The main result is the new distinction between probabilistic predictability (when the committor function is smooth and probability can be computed which does not depend sensitively on the initial condition) and probabilistic unpredictability (when the committor function depends sensitively on the initial condition).
  We also demonstrate that the Jin and Timmermann model might be the first example of a stochastic differential equation with weak noise for which transition between attractors do not follow the Arrhenius law, which is expected based on large deviation theory and generic hypothesis.
}

\begin{document}

\maketitle

\section*{Significance Statement}

A key problem for atmospheric and climate phenomena is to predict events beyond the timescale over which deterministic weather forecast is possible. \textcolor{black}{In a simple model of El Nino, we demonstrate the existence of two regimes, depending on initial conditions}. For initial conditions in the “probabilistic predictability” regime, the system is unpredictable deterministically because of chaos, but the probability of occurrence of the event can still be predicted because it depends only weakly on the initial condition. In the “probabilistic unpredictability” regime, even predicting probabilities is difficult, because the probability depends strongly on initial conditions. These new concepts of probabilistic predictability and unpredictability should be key in understanding the predictability potential for rare events in climate problems, and in other complex dynamics.

\section{Introduction}\label{sec:intro}

It has long become clear that statistics and probability are the natural languages for climate: for given boundary conditions, there is a typical state (or several, in case of bimodality), the \emph{climatology}, and fluctuations around typical conditions, referred to as \emph{climate variability}, involving various time and space scales.
At first sight, this kind of description may seem orthogonal to the problem of weather forecasting, which consists in predicting the exact state of the atmosphere at a given future time.
However, notwithstanding the use of probabilities in numerical weather forecasting for uncertainty quantification, these two approaches meet in several areas of current climate research~\citep{kalnay2003atmospheric, dijkstra2013nonlinear, ragone2018computation}.
For instance, we are often interested in predicting the occurrence of specific fluctuations of the climate system, be it a given mode of climate variability, such as the \ac{enso}~\citep{philander1990nino}, regime changes~\citep{tantet2015early},
or rare events such as heat waves~\citep{ragone2018computation}, sudden stratospheric warming, cold spells, extreme precipitations, or any other event of importance.
All these events have a probability of occurring any given year, i.e.\ with respect to climatological conditions, but one may also be interested in their probability of occurrence conditioned on the state of the climate system at the time of the prediction.
For instance, given the global impact of events like \ac{enso}, much efforts have focused on developing methods to forecast it several months in advance~\citep[e.g.][]{latif1994review, clarke2008introduction, Chekroun2011a, ludescher2014very, feng2017climate, nooteboom2018using}.
Similarly, one may want to estimate the probability of occurrence of a summer drought based on soil moisture in the spring, the probability of occurrence of a heat wave a few weeks in advance, based on the observed atmospheric circulation, or the probability of an extreme hurricane season, based on sea surface temperature.
Such forecasts are extremely challenging, but would be rewarded with proportionally large benefits, given the socio-economic impact of these events at the local and global scales, especially in a climate change context~\citep{aghakouchak2012extremes, coumou2012decade, field2012managing, herring2014explaining}.
While it is not clear that this may be reliably achieved for all the above exemples, due to their different physical nature, conceptually all these events fall in the same class of prediction problems.
The goal of this paper is to discuss the mathematical structure of such climate prediction problems.

Indeed, the mathematical structure of weather forecasting is quite clear: it consists in solving an \ac{ivp}.
Given an initial condition $x_0$ belonging to an appropriate phase space, we are interested in computing the trajectory $x_t=\phi^t x_0$, where $\phi^t$ is the flow of the dynamical system, encapsulating the evolution equations.
For many dynamical systems, this description only holds for a finite time in practice, due to sensitive dependence on initial conditions.
This limitation was already known from mathematicians in the 19th century, such as Poincaré and Hadamard.
For low-dimensional chaotic dynamical systems, this time scale, up to which deterministic forecasts are relevant, corresponds the the \emph{Lyapunov time}~\citep{castiglione2008chaos}.
In the atmosphere, the predictability horizon, about two weeks in practice, corresponds to the time it takes for undetectable errors at the smallest scales of the flow to contaminate the large scales~\citep{Thompson1957, Novikov1959, Lorenz1969a}.
Early numerical weather prediction attempts fell short of this predictability horizon, both due to model errors and sparsely constrained initial conditions.
As models improved and observational data became much denser, owing in particular to the advent of satellite observations, performance rose and skillfull forecasts are now close to the theoretical barrier~\citep{Bauer2015}.
Beyond this limit, the dynamics becomes effectively stochastic.
Notwithstanding the fact that the relevant phase space may be different for climate dynamics over geological time scales, climate therefore corresponds to the statistical properties of some stochastic process ${(x_t)}_{t>0}$.
Over very long times, we expect those statistical properties to be independent of the initial condition.
In other words, the mathematical concept relevant for climate is the \emph{invariant measure} of the system.
For lack of better techniques, in practice we still compute these properties by averaging over long times and over realizations using ensembles of trajectories obtained by numerical integration of climate models.
In any case, the invariant measure only describes the system for times larger than the \emph{mixing time}, after which the initial condition is forgotten.
However, in the applications cited above, the time scale of interest is the intermediate case for which a deterministic forecast is not relevant, but for which some information, more precise than the climate average, might be predicted.
We call this range of time scales the \emph{predictability margin}.
\textcolor{black}{We stress that the concept of predictability margin depends on the particular phenomenon under investigation.}

\begin{figure}[ht]
  \centering
  \includegraphics[width=\linewidth]{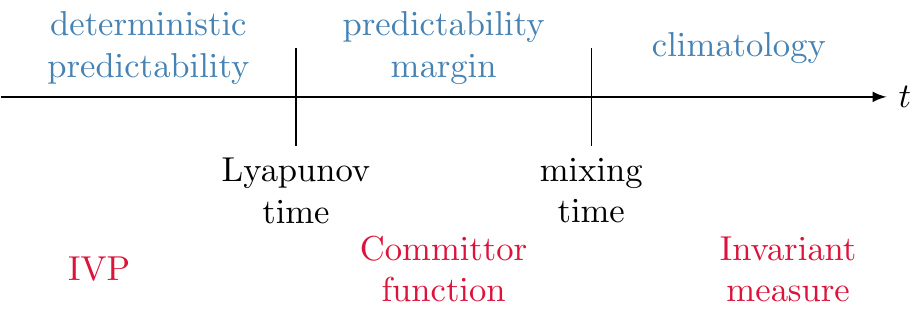}
  \caption{\label{fig:schematic} Schematic illustrating the concept of \emph{predictability margin}: deterministic predictability is only possible until a finite time (e.g.\ the Lyapunov time). The associated mathematical problem is an \acf{ivp}. Long term statistical properties (beyond the mixing time) do not depend on the initial condition, and the corresponding mathematical object is the \emph{invariant measure}. In the intermediate range of timescales, which we call the \emph{predictability margin} here, the appropriate mathematical concept is the \emph{committor function}, which encodes the probability of a given event to occur, condition on the state of the climate system at the time of the prediction.}
\end{figure}

Prediction problems at the predictability margin are of a probabilistic nature, because they are concerned with time scales beyond the deterministic predictability horizon of the system (e.g.\ the Lyapunov time).
However, we stress that the Lyapunov time scale, a global quantity, is clearly not the relevant dynamical quantity for this predictability problem.
\textcolor{black}{By contrast, at the predictability margin, the predictability depends on the current state of the system, by definition.}
Then, the question is: what is the relevant mathematical concept for prediction problems at the predictability margin? The relevant mathematical concept is called the \emph{committor function}~\citep{E2005transition, vanden2006transition}. This is a very generic concept: a committor function is the probability for an event to occur in the future, as a function of the current state of the system. Committor functions have first been introduced in climate sciences in~\citep{Lucente2019}, and have been used to study sudden stratospheric warming~\citep{finkel2020path,finkel2021learning} and to understand the flow of ocean debris~\citep{miron2021transition}. The interest of putting a name, the committor function, to this otherwise very common and generic concept, is two-fold. First it allows to study its mathematical properties and to relate them to key concepts in dynamical systems, for instance the predictability margin, as we do in the present work. Second, it comes with specific theoretical and computational approaches to compute this important quantity. \textcolor{black}{For instance, the committor function is one of the essential components of transition path theory, which is a mathematical framework for studying rare events}, see for example~\cite{vanden2006transition,metzner2006illustration,metzner2009transition} and references therein. In atmosphere dynamics, a very interesting use of smart ways of estimating the committor function for a simplified model of sudden stratospheric warming is provided by~\cite{finkel2021learning}.

Many problems in \emph{medium-range forecasting} fall within the \emph{predictability margin} range; to illustrate the interest of committor functions, we will select only one example of application, the problem of \ac{enso} prediction, using a very simple model.
While, as mentioned above, many studies strive to predict the occurrence of El Niño a few months in advance, we shall address here a slightly different problem, focusing on predicting strong El Niño events on longer time scales.
This is also a relevant question from the point of view of climate dynamics; while strong El Niño events have been observed almost periodically since the 1950s, with a return time around 15--20 years, historical data and paleoclimatic proxies indicate that \ac{enso} may exhibit high variability over centennial timescales~\citep{Cobb2003, Khider2011, McGregor2013} and beyond~\citep{Rickaby2005, Fedorov2006, Cobb2013}.
We study the dynamics of a low-dimensional stochastic model proposed to explain the decadal amplitude changes of \ac{enso}, the Jin and Timmermann model~\citep{timmermann2002nonlinear, timmermann2003nonlinear}.
This model is not aimed at reproducing any precise properties of the real El Niño Southern Oscillation.
It is rather used as a paradigm example to introduce the concept of a committor function, and to study its main properties.
This will lead us to define probabilistic predictability and unpredictability, some concepts that should be useful for other applications.

We show that probabilistic prediction at the predictability margin depends on the initial state, and that probabilistic predictability is encapsulated in the committor function.
This property is analogous to classical, deterministic predictability, which is known to depend on the state of the system: some circulation patterns, such as the positive phase of the \ac{nao}, lead to improved predictability.
However, we stress that deterministic and probabilistic predictability are different concepts: probabilistic predictability means that the probability of the event does not depend sensitively on the initial conditions.
This corresponds to a region of phase space where the committor function has gentle variations with the initial conditions.
In these areas, the event occurs with a probability $p$ that can be easily determined in practice because of these gentle variations.
On the contrary, probabilistic unpredictability corresponds to regions of the phase space with a rough committor function.
In these regions, the occurence of the event is also probabilistic.
But the probability itself has very rapid variations with the initial conditions, which make the prediction highly dependent on the level of precision with which the initial condition is known.
The existence of such features, and especially the new and most interesting \textit{probabilistically predictable} region, should be generic for most prediction problems in climate dynamics.

This paper also discusses relations between qualitative properties of the committor function, finite time Lyapunov exponents, and the stability properties of trajectories with respect to noise perturbations.
It also discuss methodological aspects for a data-based approach for the computation of committor functions.

The dynamics of the Jin and Timmermann model, when perturbed by a weak noise, is characterized by rare transitions between a limit cycle and a strange attractor~\citep{roberts2016mixed, guckenheimer2017predictability}.
Based on large deviation theory, and with generic hypothesis, the average transition time $\mathbb{E}[\tau_c]$ to see such transitions is expected to change following an Arrhenius law: $\mathbb{E}[\tau_c] \underset{\sigma \rightarrow 0}{\asymp} A\exp\left(\Delta V/\sigma^2\right)$, where $\sigma$ is the noise amplitude.
Using numerical simulations, we demonstrate that the Jin and Timmermann transition times do not follow the expected Arrhenius law for a very large range of small noise amplitudes.
We conjecture that this very interesting phenomenon might be the first observed counterexample to the expected generic result, for deterministic dynamics perturbed by weak noises.
We argue that this is related to the intricated entanglement between the basins of attraction of the limit cycle and the strange attractor.

The paper is organized as follows: in Sec.~\ref{sec:jintimmermann} we define the Jin and Timmermann model~\citep{timmermann2002nonlinear, timmermann2003nonlinear}.
In this model, the occurrences of strong \ac{enso} events correspond to noise-induced transitions between a strange attractor and a limit cycle~\citep{roberts2016mixed, guckenheimer2017predictability}.
We study in Sec.~\ref{sec:firstpassage} the statistics of such transitions, and we show that they do not obey an Arrhenius law.
Finally, in Sec.~\ref{sec:committorenso} we introduce the committor function, we compute it for the Jin-Timmermann model, and characterize the regions of the phase space with qualitatively different predictability properties.
In the regime of intermediate noise amplitude, at the predictability margin, we delineate four regions (see Fig.~\ref{X111}): two regions of \emph{deterministic predictability} (where the event occurs with probability $0$ or $1$), one \textit{probabilistically predictable} region (where a value of the probability $0<q<1$ can clearly be predicted with very mild dependence with respect to initial condition), and finally a region which is unpredictable in practice, because the strong dependence with respect to the initial condition prevents any precise prediction, either deterministic or probabilistic.

\section{The Jin and Timmermann model}\label{sec:jintimmermann}
\acf{enso} is one of the most important mode of climate variability at the interannual time scales~\citep{philander1990nino}.
El Ni\~{n}o events consist in an increase of the Sea Surface Temperature in the eastern equatorial Pacific Ocean, leading at the local scale to reduced thermocline depth, reduced upwelling and reduced nutrient supply, thereby affecting marine life.
Such events are correlated with a reorganization of the Walker circulation in the atmosphere, known as the Southern Oscillation.
The global phenomenon, referred to as \ac{enso}, has major impacts all over the world.
However, the nonlinear coupled atmosphere-ocean dynamics of \ac{enso} makes it very difficult to predict~\citep[e.g.][]{mcphaden2015curious}.
Models of various complexities have been constructed to capture the dynamics of El Ni\~{n}o at different levels of realism~\citep{clarke2008introduction, sarachik2010nino}.
In order to introduce and illustrate the concept of committor function in the simplest possible framework, we shall consider here one of the most idealized models, consisting of a low-dimensional stochastic process.
This simple dynamical model, introduced by~\cite{jin1997equatorial1,jin1997equatorial2}, accounts for the recharge-discharge mechanism which is at the basis of \ac{enso}.
This model was later extended by~\cite{timmermann2003nonlinear} and was related to the decadal amplitude changes of \ac{enso}~\citep{timmermann2002nonlinear}.
The model describes the evolution of three variables:
\begin{enumerate}
\item $T_1$, the Sea Surface Temperature in the western equatorial Pacific Ocean,
\item $T_2$, the Sea Surface Temperature in the eastern equatorial Pacific Ocean,
\item $h_1$, the thermocline depth anomaly in the western Pacific.
\end{enumerate}
Assuming a thermal relaxation towards a radiative-convective temperature $T_r$, the equations of motion can be written as~\citep{timmermann2002nonlinear,timmermann2003nonlinear}:
\begin{align}&\frac{{\rm d}T_1}{{\rm d} t}=-\alpha(T_1-T_r)-\epsilon\beta\tau(1-\sigma \eta_t)(T_2-T_1), \nonumber\\
&\frac{{\rm d} T_2}{{\rm d} t}=-\alpha(T_2-T_r)+\zeta\beta\tau(1-\sigma \eta_t)(T_2-T_{sub}), \nonumber\\
&\frac{{\rm d} h_1}{{\rm d} t}=r\left(-h_1-\frac{1}{2}bL\tau\right),
\label{Eq:JTDimension}
\end{align}
where $\epsilon$ and $\zeta$ represent the strength of the zonal and vertical advection, $T_{sub}$ denotes the temperature being upwelled into the mixed layer, $\tau$ represents the zonal wind stress, $L$ denotes the basin width, $b$ captures the efficiency of wind stress in driving thermocline tilt, $1/\alpha$ measures a typical thermal damping timescale and $1/r$ is the dynamical adjustment timescale of the thermocline depth.
The term $\eta_t$ in the equations for temperatures is a Gaussian white noise with unit variance and the level of stochasticity is controlled by the noise amplitude $\sigma$.
This term takes into account the fluctuating component of wind stress.
In the last equation the noise has not been considered because wave processes are filtered out in the thermocline equations of the model~\citep{timmermann2003nonlinear}.

The expressions of $T_{sub}$ and $\tau$ are
\begin{align}
&T_{sub}=\frac{T_r+T_{r_0}}{2}+\frac{T_r-T_{r_0}}{2}\tanh{\left(\frac{H+h_2-z_0}{h^*}\right)}, \nonumber\\
		   &\tau=\frac{\mu(T_2-T_1)}{\beta},
\end{align}
where $T_{r_0}$ is a reference temperature, $h_2$ is the thermocline departure from its reference value $H$, $z_0$ represents the depth at which $\zeta$ takes its characteristic value, $h^*$ measures the sharpness of the thermocline.
The relation between the eastern and western thermocline depth anomalies is
\begin{equation}
  h_2=h_1+bL\tau.
\end{equation}

In order to study the dynamical behavior of the system it is useful to perform a change of variables from physical to dimensionless ones~\citep{roberts2016mixed}.
So, we define
\begin{align*}
  &x=\frac{T_2-T_1}{T_0}\quad,\quad y=\frac{T_1-T_r}{T_0},\\
  &z=\frac{h_1+H-z_0}{h^*} \quad , \quad \tilde{t}=\frac{t}{t^*},
\end{align*}
where $T_0=\frac{h^* \beta}{bL\mu}$ and $t^*=\frac{bL}{\beta\zeta h^*}$.
After the change of variables, the equations~\eqref{Eq:JTDimension} read
\begin{align}
&\dot{x}=\rho\delta(x^2-ax)+x\lbrack x+y+c-c\tanh{(x+z)}\rbrack-D_x(x,y,z) \eta_{t}, \nonumber \\
&\dot{y}=-\rho\delta(x^2+ay)+D_y(x,y,z)\eta_{t}, \nonumber \\
&\dot{z}=\delta\left(k-z-\frac{x}{2}\right),
\label{eq:JTdimensionless}
\end{align}
where
\begin{align*}
  D_x(x,y,z)&=[(1+\rho\delta)x^2+xy+cx(1-\tanh{(x+z)})]\sigma,\\
  D_y(x,y,z)&=\rho\delta x^2\sigma,
\end{align*}
and the new control parameters $\delta,\rho,c,k$, and $a$ are defined as follows:
\begin{align*}
  &\delta=\frac{rbL}{\zeta\beta h^*},& \rho&=\frac{\epsilon h^* \beta}{rbL},&&\\
  &a=\frac{\alpha b L}{\epsilon \beta h^*},& c&=\frac{T_r-T_{r_0}}{2T_0},& k&=\frac{H-z_0}{h^*}.
\end{align*}

The deterministic version ($\sigma=0$) of equations~\eqref{eq:JTdimensionless} was widely studied in literature.
For some parameter values, the system has only one attractor, a periodic orbit.
Figure~\ref{ComparisonModelNinoIndex}-b illustrates such a periodic orbit, with the parameter values {$[\delta,\rho,c,k,a]=[0.2625,0.3224,2.3952,0.4032,6.8927]$ and dimensional normalization constants $[T_0, t^*, h^*] = [2.8182\, ^\circ \text{C}, 104.9819\, \text{days}, 62 \, \text{m}]$.
\cite{roberts2016mixed} also analyzed the mechanism through which this limit cycle arises.
\cite{roberts2016mixed} defined strong El-Ni\~{n}o events for this model as periods in this limit cycle for which the temperature is large.
Figure~\ref{ComparisonModelNinoIndex} shows a qualitative comparison of the eastern Pacific sea surface temperature anomaly for this limit cycle with the El-Ni\~{n}o3 index.
Both the measurements and the model display positive temperature anomaly excursions with a return time of approximately $15$ years.

\begin{figure}[ht]
\centering
\includegraphics[width=0.8\linewidth]{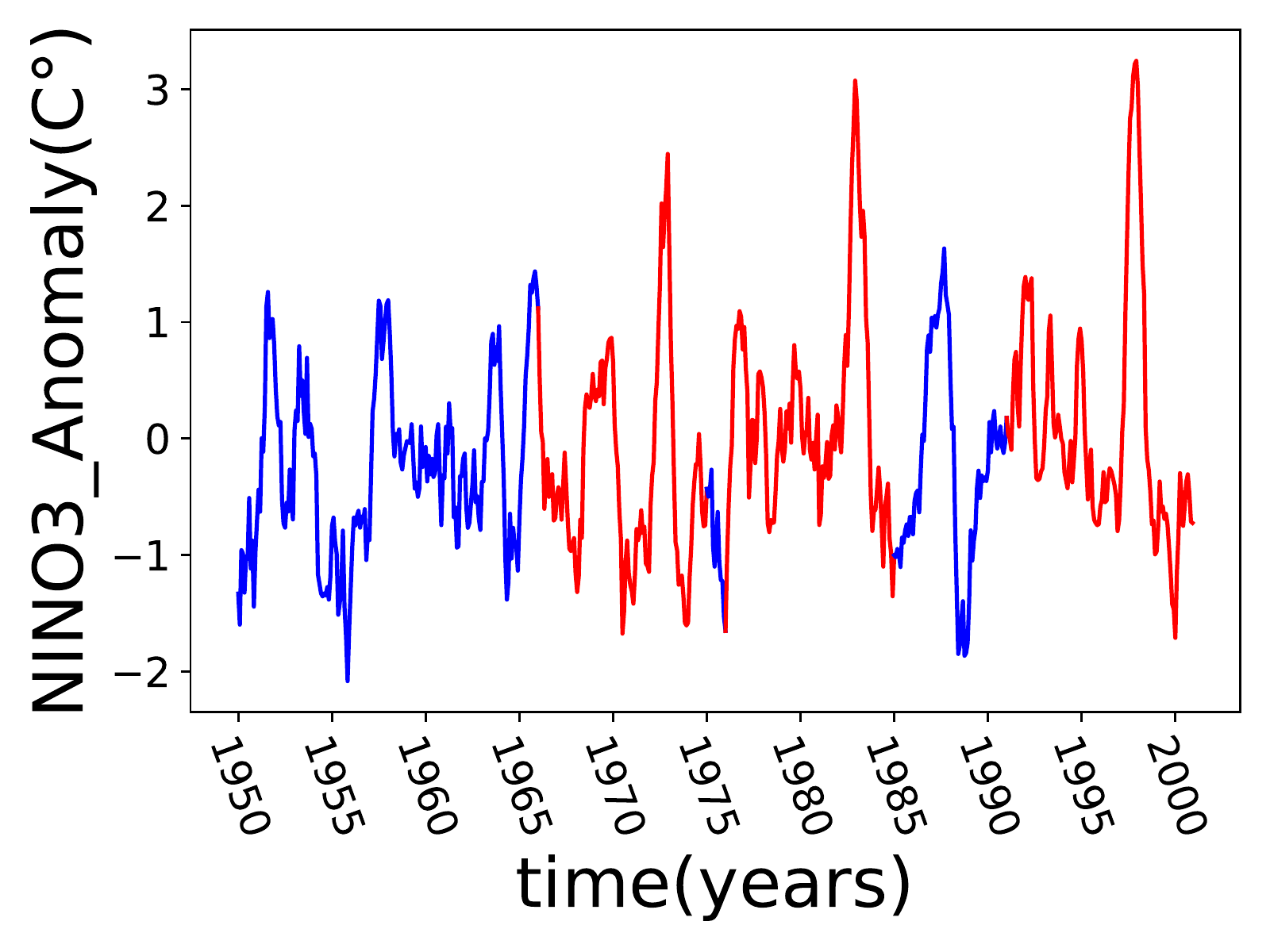}
\includegraphics[width=0.8\linewidth]{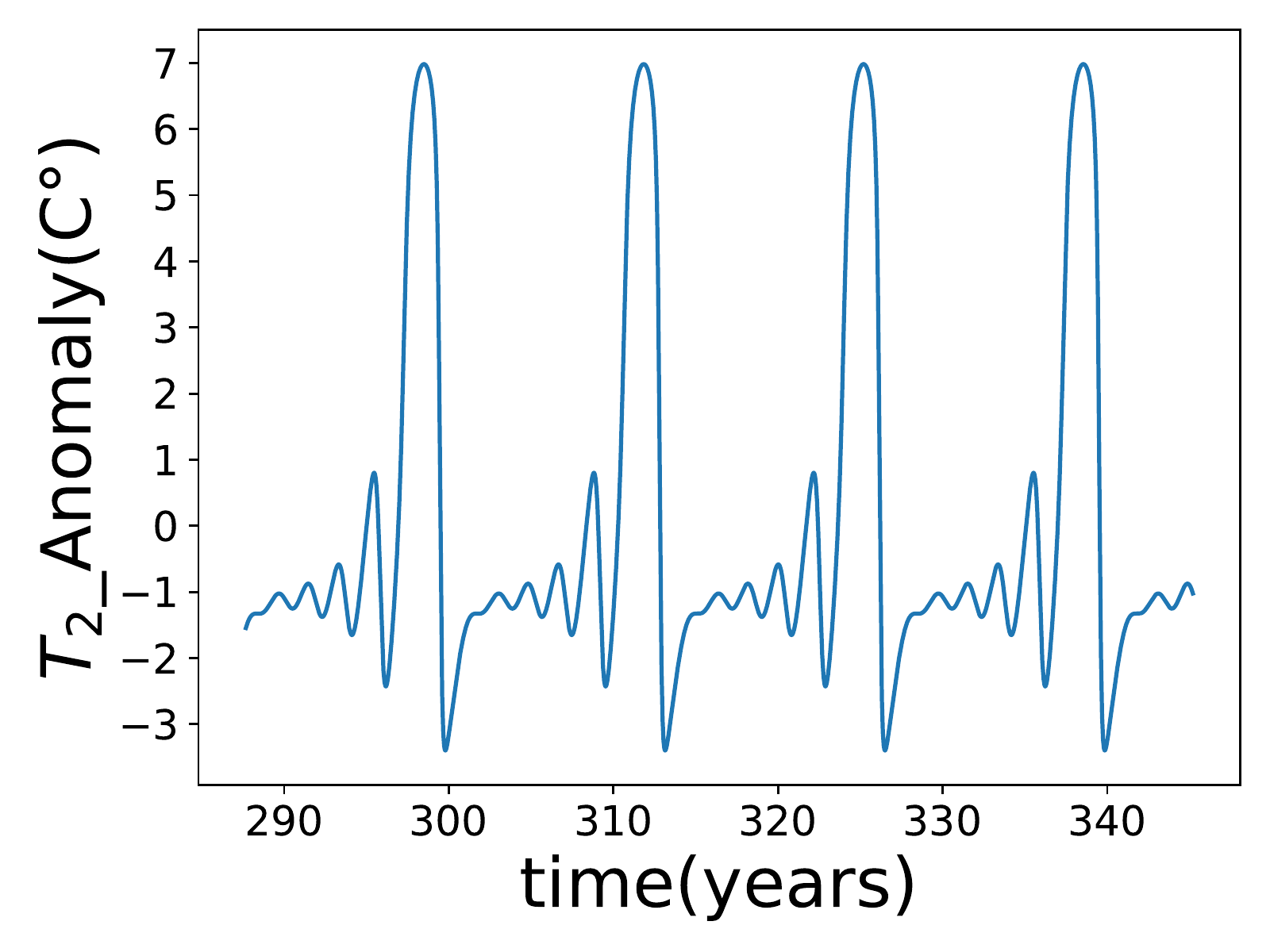}
\caption{Top plot: observed sea surface temperature anomalies over the last decades, spatially averaged over the Ni\~{n}o-3 region. Bottom: eastern Pacific sea surface temperature anomalies simulated with the deterministic Jin and Timmermann model. \textcolor{black}{The red segments in the upper panel highlight the growing sequence of oscillations leading to a strong El-Ni\~no event, which are qualitatively reflected in the behavior of the model shown in the lower panel.}}
\label{ComparisonModelNinoIndex}
\end{figure}

Varying the parameter $\delta$, a strange attractor emerges through a period doubling cascade, as shown by~\cite{guckenheimer2017predictability}.
\textcolor{black}{Moreover,~\cite{guckenheimer2017predictability} show that for some parameter values this strange attractor coexists with a limit cycle similar to the one mentioned in the previous paragraph (with a different period).}
Following~\cite{guckenheimer2017predictability}, we use the parameter values $[\delta,\rho,c,k,a]=[0.225423,0.3224,2.3952,0.4032,7.3939]$ all along this paper.
While~\cite{guckenheimer2017predictability} considered only the deterministic model ($\sigma=0$), we also consider later on the stochastic model ($\sigma\neq0$).
For $\sigma=0$, the strange attractor and the limit cycles are shown in Fig.~\ref{TwoAttractors}.
They are intertwined in a complex way.

For this dynamics, we define a strong El-Ni\~{n}o event as any situation when $x$ becomes larger than the threshold $x_c=-1$.
As can be seen from Fig.~\ref{TwoAttractors}, this only happens in the limit cycle, and not in the strange attractor.
\begin{figure}[ht]
\centering
\includegraphics[width=\linewidth]{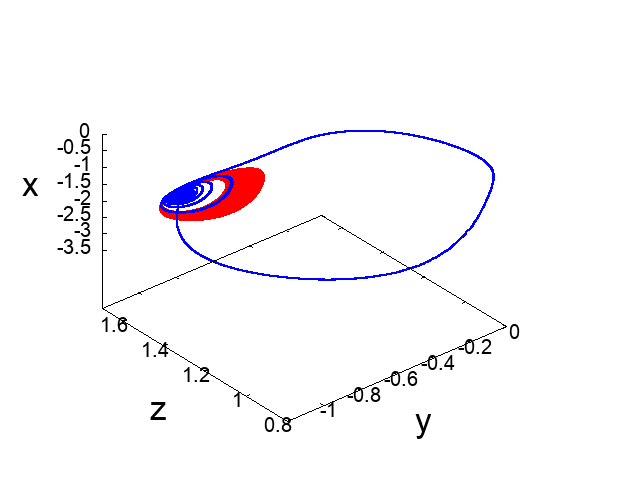}
\caption{The two intertwined attractors of the Jin-Timmermann model (limit cycle in blue and strange attractor in red).}
\label{TwoAttractors}
\end{figure}
Note that with this choice of parameters, the return period of strong El Niño events on the limit cycle is around 50 years (the time unit is $t^*=105$ days and the period is $186$ non-dimensional time units), rather than 15 for the parameters studied by~\cite{roberts2016mixed}.

For $\sigma=0$, when the parameter $a$ is time periodic rather than constant, mimicking a seasonal forcing,~\citep{guckenheimer2017predictability} observed transitions between the strange attractor and the limit cycle.
In the following we consider a constant $a$, and rather study noise induced transitions between these two attractors, for $\sigma \neq 0$.

We now discuss qualitatively the effect of the noise level $\sigma$.
For small $\sigma>0$, the dynamics can switch from one regime close to the strange attractor to another regime close to the limit cycle.
\begin{figure}[ht]
\centering
\includegraphics[scale=0.5]{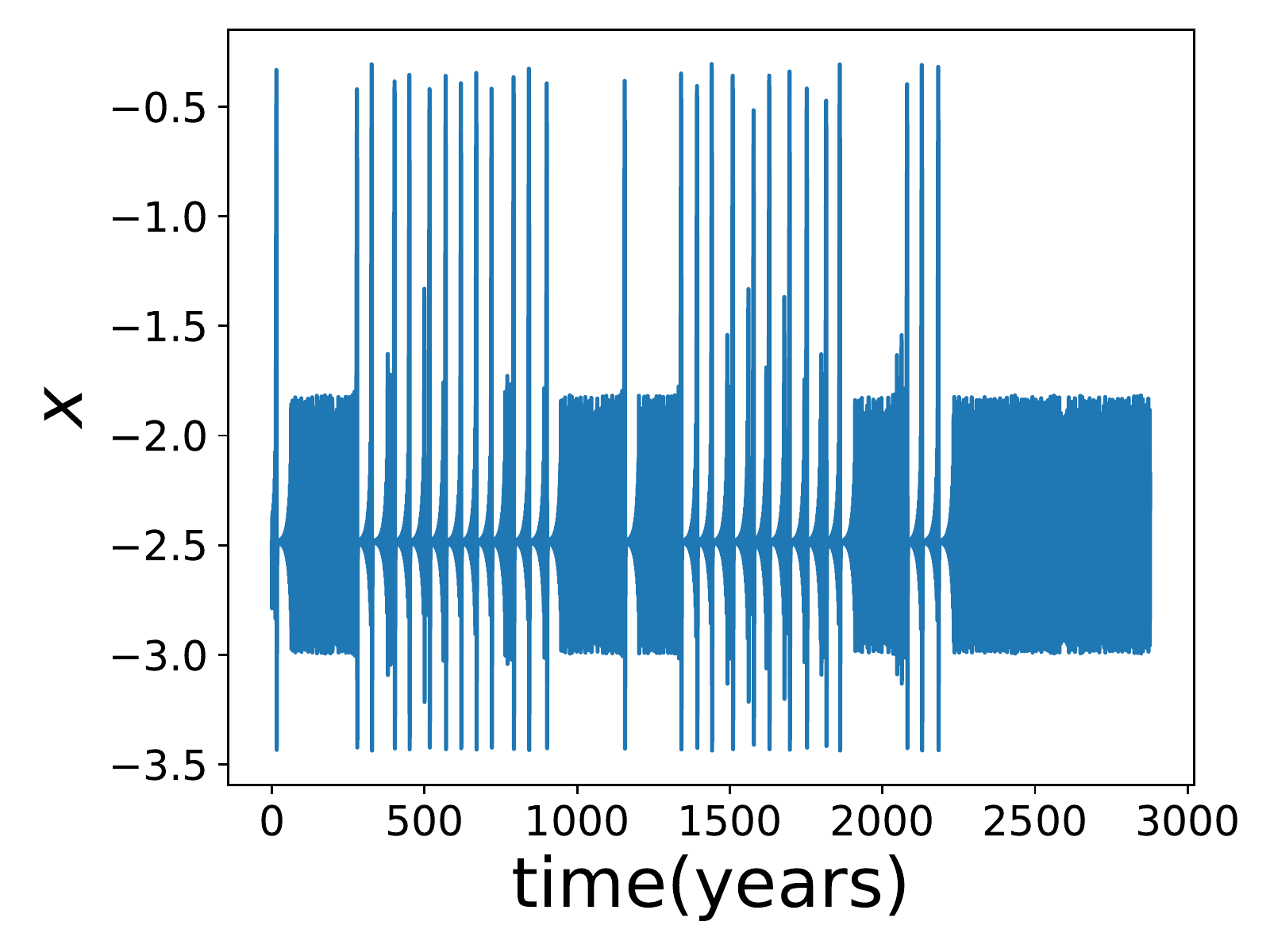}
\caption{Time series of the variable $x$ for $\sigma=0.00005$. The dynamics exhibits a switching between a regime close to the strange attractor (bounded regions around $x=-2.5$) and a limit cycle regime.}
\label{Fig:ENSODynamics}
\end{figure}
This is evident by looking at Fig.~\ref{Fig:ENSODynamics}, where the time evolution of the variable $x$ for $\sigma=5\cdot10^{-5}$ is shown.
As can be seen, the system switches between epochs where the variable $x$ is bounded around the value $x=-2.5$ (strange attractor regime) and epochs where there are large oscillations with a period around $50$ years (limit cycle regime).
\textcolor{black}{Such shifts can be interpreted as a toy model for hypothesized regime changes in the long-term ENSO dynamics.}
Strictly speaking, for $\sigma \neq 0$, there is not anymore two attractors.
However, as illustrated in Fig.~\ref{Fig:ENSODynamics}, for small enough $\sigma$, the trace of the two deterministic attractors is clearly visible.
For small $\sigma$, we will thus continue to discuss the strange attractor and the limit cycle, for simplicity.
Strong El-Ni\~{n}o events occur only during the limit cycle regime.
In the next section, we will study transitions from the strange attractor regime to the limit cycle regime (or equivalently to the strong El Ni\~{n}o regime).

The frequency of transitions from the strange attractor to the strong El-Ni\~no regime increases as the amplitude of the noise increases.
For large values of $\sigma$, the dynamics is completely dominated by the noise and the distinction between the two attractors becomes meaningless.

\section{Statistics of the first exit times for transitions to strong El Ni\~{n}o regimes}\label{sec:firstpassage}

As discussed in the previous section, we define strong El Ni\~{n}o events as periods of time when $x_c>-1$, which occur along the limit cycle.
In this section, we study transitions from the strange attractor regime to the strong El Ni\~{n}o regime, and how their statistics change when the noise amplitude $\sigma$ is varied.

We consider $\mathbf{X(t)}=(x(t),y(t),z(t))$ solutions to the stochastic Jin and Timmermann model~\eqref{eq:JTdimensionless}.
We define \emph{first exit times} from a point $\mathbf{x}$ to the strong El Ni\~{n}o regime as
\begin{equation}
\tau_c(\mathbf{x}) = \inf \{ t>0: x(t)>x_c\ |\ \mathbf{X}(0)=\mathbf{x} \}. \label{eq:firstpassagetime}
\end{equation}
The random variable $\tau_c(\mathbf{x})$ depends both on the realization of the noise and on the initial condition $\mathbf{x}$.
The statistics are understood as averages over both the noise realization and the invariant measure of $\mathbf{x}$ over the strange attractor of the deterministic system ($\sigma=0$), the so called SRB measure.
For instance the mean first exit time $\mathbb{E}[\tau_c]$ is defined as
\begin{equation}
\mathbb{E}[\tau_c] = \int{\rm{d}\mathbf{x}\,\rho_{SRB}(\mathbf{x})\mathbb{E}_{noise}[\tau_c(\mathbf{x})]}.
\label{eq:AverageFirstPassageTime}
\end{equation}
where $\mathbb{E}_{noise}[\cdot]$ is the expectation with respect to the noise realization and $\rm{d}\mathbf{x}\,\rho_{SRB}(\mathbf{x})$ is the SRB measure.

The SRB measure is defined through time averages of the deterministic dynamics ($\sigma=0$).
In practice, we thus compute a very long trajectory of the deterministic dynamics.
We then choose a set of 1000 initial conditions $\mathbf{x}$ taken randomly among all the points of this deterministic trajectory.
Then, for any fixed value of $\sigma >0$, for any initial condition  $\mathbf{x}$, we compute the first-passage time $\tau_c$ for several noise realizations.

In Fig.~\ref{fig:fptpdf}, we show the probability density function $p(\tau_c)$ of $\tau_c$ based on this ensemble.
\begin{figure}[ht]
  \centering
  \includegraphics[width=\linewidth]{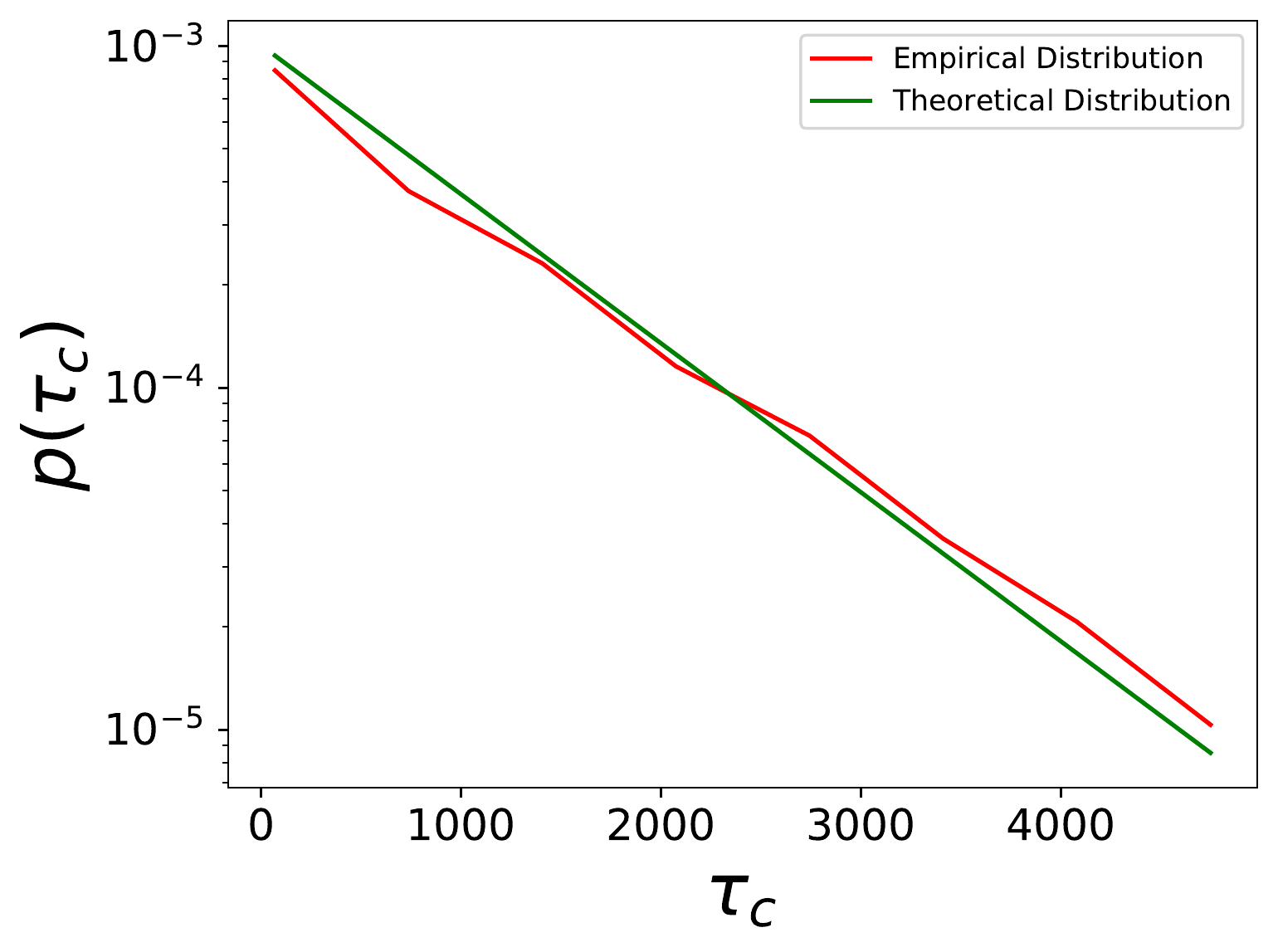}
  \caption{\label{fig:fptpdf} Logarithm of the Probability Density Function of the first exit time between the strange attractor and the limit cycle for $\sigma=5\times 10^{-5}$ sampled by direct integration (red) of the stochastic differential equation~\eqref{eq:JTdimensionless}, and the exponential distribution $p(\tau_c)=\lambda e^{-\lambda\tau_c}$ with $\lambda^{-1} = \mathbb{E}[\tau_c]$ (green).}
\end{figure}
The probability density function is close to an exponential: $p(\tau_c)=\lambda e^{-\lambda\tau_c}$.
The parameter $\lambda$ is then equal to the inverse of the mean first exit time: $\lambda^{-1} = \mathbb{E}[\tau_c]$.

Because typically  $\tau_c(\mathbf{x})$ is much larger than the relaxation time to the strange attractor, one might expect that for most of the points of the strange attractor the dependence of $\tau_c(\mathbf{x})$ on $\mathbf{x}$ is practically irrelevant.
Indeed, we have verified numerically that except for a small region around the transition paths, the statistics are independent from the initial condition, up to numerical accuracy.
Hence we have $\mathbb{E}[\tau_c] \simeq \mathbb{E}_{SRB}[\tau_c(\mathbf{x})] \simeq \mathbb{E}_{noise}[\tau_c(\mathbf{x})]$ for generic points $\mathbf{x}$ close to the strange attractors.

As illustrated in Fig.~\ref{fig:fptpdf}, the mean first exit time $\mathbb{E}[\tau_c]$ is of the order of 1,000 in non-dimensional time units.
The measured value is $\tau_c=1,039$.
As $\tau_c$ is much larger than the mixing time of the SRB measure, of order 1, then it is natural to expect that the first exit times should be random and distributed, with a very good approximation, according to a Poisson statistics.
The observed exponential distribution is consistent with such a Poisson statistics.
Similar exponential distributions for first exit times were observed for the deterministic dynamics with periodic modulation of the $a$ coefficient~\citep{guckenheimer2017predictability}.

We now study how the mean first exit time $\mathbb{E}[\tau_c]$ varies when the noise amplitude $\sigma$ is changed.
One generally expects an Arrhenius law:
\begin{equation}
  \mathbb{E}[\tau_c] \underset{\sigma \rightarrow 0}{\sim} Ae^{\frac{\Delta V}{\sigma^2}}.
  \label{Eq:Arrhenius}
\end{equation}
Arrhenius laws where first derived by Kramers for gradient dynamics forced by white noise $\dot{\mathbf{x}} = - {\rm d}V/{\rm d}\mathbf{x} + \sqrt{2\sigma}\eta(t)$, where $\Delta V$ (the \emph{potential barrier}) is the difference of potential between the original attractor and the saddle-point separating the basins of attraction of the two attractors (see for example the textbook by~\cite{gardiner1985handbook}).
The Jin and Timmermann model is however not a gradient dynamics, and the function $V$ is not explicit.
For such non gradient systems, the exponential factors of the Arrhenius law can be justified through a Laplace principle for a path integral representation of the transition probabilities, or asymptotic studies of Fokker-Planck operators~\citep{Graham1987macroscopic}, or through large deviation theory~\citep{Freidlin_Wentzell}.
The function $V$ is then called the \emph{quasipotential}, which can be computed through a variational problem, or computing viscosity solutions of a Hamilton-Jacobi equation.
The sub-exponential prefactor $A$ in Eq.~(\ref{Eq:Arrhenius}) can be computed through Eyring-Kramers formulas, derived either for gradient~\citep{bovier2004metastability} or non-gradient dynamics~\citep{Bouchet_Reygner_2016}, for transitions from a point attractor and through a point saddle.
Many generalizations exist, for instance for periodically modulated systems~\citep{dykman2005activated} or systems approaching a bifurcation~\citep{Herbert2017}.
In large dimensional systems related to climate dynamics, effective Arrhenius laws have been observed numerically, for instance in transitions in beta-plane turbulence~\citep{Bouchet2019} or in a simplified climate model with ice-albedo feedback~\citep{lucarini2019transitions}.

When one studies a transition from a strange attractor rather than a point attractor, or when the saddle set between the two basins of attraction is a not a simple saddle point, then no theorem exists to put the Arrhenius law (\ref{Eq:Arrhenius}) on firm mathematical ground.
However it has been argued for a long time~\citep{Graham1987macroscopic}, that if a finite distance $d>0$ exists between the strange attractor and the saddle set, then there is a non-zero quasipotential difference $\Delta V>0$, and an Arrhenius law should generically be expected.

In Fig.~\ref{MeanExitTime}, we show the mean first exit time $\mathbb{E}[\tau_c]$ as a function of the noise amplitude $\sigma$.
\begin{figure}[ht]
\centering
\includegraphics[scale=0.5]{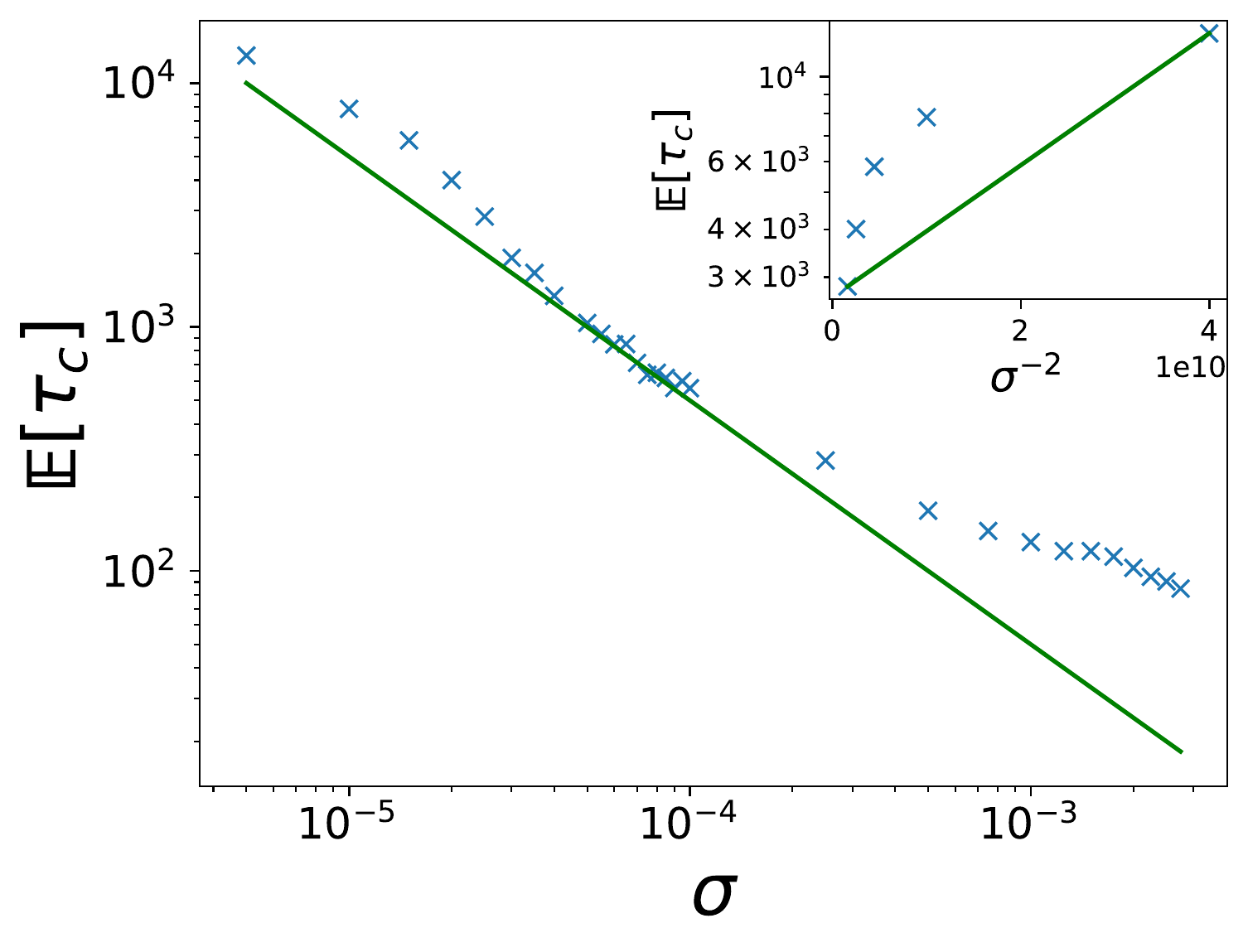}
\caption{Mean first exit time $\mathbb{E}[\tau_c]$ for the transition from the strange attractor regime to the strong Ni\~{n}o event regime, as a function of the noise amplitude $\sigma$, in log-log coordinates. In the limit of small noise amplitude $\sigma < 10^{-4}$, $\mathbb{E}[\tau_c]$ seems to be closer to a power-law $\sigma^{-1}$ (green line) than to the standard Arrhenius law. \textcolor{black}{The inset shows the mean first exit time $\mathbb{E}[\tau_c]$ as a function of $\sigma^{-2}$ for small values of $\sigma$. This clearly shows that $\mathbb{E}[\tau_c]$ does not follow an Arrhenius law (represented by the green line).}}
\label{MeanExitTime}
\end{figure}
It ranges from transition times of about 25 years ($T=100$ in non-dimensional time units, for the strongest values of the noise amplitude) to several millenia (about 3000 years for $T=10 000$ in non-dimensional units for weak noise).
Fig.~\ref{MeanExitTime} clearly shows that the mean exit time from the strange attractor to the regime of strong Ni\~{n}o events does not follow an Arrhenius law of the form~\eqref{Eq:Arrhenius}.
The mean exit time seems much closer to a power law $\mathbb{E}[\tau_c]\propto \sigma^{-1}$. 
\textcolor{black}{An even clearer visualization of the violation of the Arrhenius law is provided in the inset of Fig.~\ref{MeanExitTime} where $\mathbb{E}[\tau_c]$ is shown as a function of $\sigma^{-2}$, for $\sigma\ll1$.}

In order to check the numerical robustness of our result, we computed $\mathbb{E}[\tau_c]$ using two different schemes of integration, and we checked numerical convergence with respect to the time step $\Delta t$ in the integration schemes.
The first integration scheme is the fourth order Runge Kutta method to which a zero mean gaussian white noise is added.
The variance of the noise is proportional to the integration time step $\Delta t$.
In this way we have a precision of $\Delta t^4 $ for the deterministic part while we make an error of order $\sqrt{\Delta t}$ for the statistics.
The second integration scheme is the stochastic Runge Kutta method which has an error of order $\Delta t$ for a stochastic dynamics~\citep{roberts2012modify}.

Let us note that many other behaviors than exponential ones have been observed for mean exit times.
For instance a power-law has been observed for flow reversals in numerical simulations of inviscid turbulent flows~\citep{shukla2016statistical}.
However for this last example, as the dynamics is not a deterministic system with attractors perturbed by weak noise, it was not clear why one should have expected an Arrhenius law in the first place.

We observe a breakdown of the Arrhenius law for the Jin and Timmermann model which is a deterministic system with attractors perturbed by weak noise.
This is striking.
Indeed we stress again that if a finite distance $d>0$ exists between the strange attractor and the saddle set, then there is a non-zero quasipotential difference $\Delta V>0$, and an Arrhenius law should be expected.
The distance $d$ might be expected to be generically strictly larger than 0.

We see two possible heuristic explanations for this interesting breakdown.
The first explanation might be that a finite distance $d>0$ and a quasipotential barrier $\Delta V>0$ between the strange attractors and the basin of attraction limit cycle do actually exist, but they are extremely small.
Then the explanation of the observed breakdown in Fig.~\ref{MeanExitTime} would be that we have not studied small enough values of $\sigma$.
We note however that we computed first exit times of order $\mathbb{E}[\tau_c] = 5.10^5$ for values of $\sigma$ as small as $10^{-6}$.
If this first explanation is valid, this means that the Arrhenius law is practically irrelevant even if it might be mathematically correct.

The second possible explanation might be that there exists no finite distance between the strange attractor and a possible fractal boundary between the basins of attractions.
Then for any small values $d$ and $v$, there always exist points in the strange attractor and in the boundary of the basin of attraction at a distance smaller than $d$ and a quasipotential differences $\Delta V$ smaller than $v$.
Many phenomenologies could then be imagined, for instance with a distribution of a large number of transition paths, possibly infinite, leading to a power law or an effective behavior of the first exit times described by any function.
Those conjectures are not based on any mathematical results yet.
However the possibility of a breakdown of the Arrhenius law is a very interesting problem, that should be studied further either through theory and mathematics, or through numerical simulations.

\section{Committor function of the Jin and Timmermann model}\label{sec:committorenso}

In Sec.~\ref{sec:firstpassage}, we have shown that, in the stochastic Jin and Timmermann model, transitions between the strange attractor regime and the strong El-Ni\~{n}o regime occur at random times, in the limit of small noise $\sigma \rightarrow 0$.
In this section, we focus on the associated prediction problem: What is the probability that a strong El Niño event occurs within a given timeframe, given the state of the system at the time of prediction?
We will address this question for any finite value of the noise amplitude $\sigma$.

We consider solutions $\mathbf{X}(t)=\left(x(t),y(t),z(t)\right)$ of the stochastic Jin an Timmermann model (\ref{eq:JTdimensionless}).
We remind the reader that we identify a strong El Niño with an event when $x >-1$.
For a solution $\mathbf{X}(t)$ that starts from $\mathbf{x}$, that is $\mathbf{X}(0)=\mathbf{x}$, we want to predict the probability $q({\mathbf x} )$ that a strong El Niño event occurs within a fixed time $T$.
This is
\begin{equation}
q({\mathbf x} )= \mathbb{P}\left(\max_{0 \leq t \leq T}\, x(t)>-1 \ |\ \mathbf{X}(0)=\mathbf{x}\right).\label{eq:Committor_x}
\end{equation}
Recalling the definition of the first passage time to a strong El Ni\~{n}o regime, Eq.~\eqref{eq:firstpassagetime},
\begin{equation}
\tau_c(\mathbf{x}) = \inf \{ t>0: x(t)>-1\ |\ \mathbf{X}(0)=\mathbf{x} \},
\end{equation}
we note that $q(\mathbf{x})=\mathbb{P}[\tau_c(\mathbf{x}) < T]$ is the \ac{cdf} of the first-passage time.
We now define committor functions and explain that $q$ is a committor function.

\textbf{Committor functions.} For a Markov stochastic process $\{\mathbf{Y}(t)\}$ which takes values in $\Gamma$, we define the \textit{first hitting time} of the set $C$ as $\tau_C(\mathbf{y})=\inf\{t: \mathbf{Y}(t) \in C\ |\ \mathbf{Y}(0)=\mathbf{y}\}$.
For two disjoint subsets $A,\, B \subset \Gamma$, the committor function $\tilde{q}(\mathbf{y})$ is defined as the probability to hit the set $B$ before hitting the set $A$:
\begin{equation}
\tilde{q}(\mathbf{y})=\mathbb{P}(\tau_B(\mathbf{y})<\tau_A(\mathbf{y})).
\label{eq:Committor}
\end{equation}

Considering the auxiliary process $\left\{\mathbf{Y}(t)\right\}$, with $\mathbf{Y}(t) =\left( \mathbf{X}(t),t \right)$, and the two sets
\begin{align}
&A=\{\mathbf{y}=(\mathbf{x},t)\,  \ |\ \,x>-1 \,  \, {\rm and} \, \, t\in[0,T]\}\, {\rm and} \nonumber \\
&B=\{\mathbf{y}=(\mathbf{x},T)\,;\, x\leq-1\},
\end{align}
we see that $q({\mathbf x} )=\tilde{q}(\mathbf{x},0)$.
Hence $q$, in Eq.~(\ref{eq:Committor_x}) is a committor function.

For an ergodic process, replacing statistical averages by temporal averages in (\ref{eq:Committor}), and using $\mathbf{y}=(\mathbf{x},t)$, we have
\begin{align}
&\rho(\mathbf{x})q(\mathbf{x})=\lim_{t\rightarrow\infty}\frac{1}{t}\int^t_{0}\text{d}t'\,\delta\left(\mathbf{X}_{t'}-\mathbf{x}\right)1_{\{\tau_{B}\leq\tau_{A}\}}, \nonumber \\
&\rm{and\ }\rho(\mathbf{x})=\lim_{t\rightarrow\infty}\frac{1}{t}\int_{0}^{t}\text{d}t'\,\delta\left(\mathbf{X}_{t'}-\mathbf{x}\right),
\label{eq:Committor_Trajectory}
\end{align}
where $\rho(\mathbf{x})$ is the stationary distribution function of $X$, $\delta$ is a Dirac delta function, and $1_{\{\tau_{B}\leq\tau_{A}\}}$ takes value $1$ if $\tau_{B}\leq\tau_{A}$ and $0$ otherwise.
The formulas (\ref{eq:Committor_Trajectory}) can be used to estimate $q(\mathbf{x})$ from an observed trajectory $\{\mathbf{X}(t)\}$ of the dynamical system.
For the sake of completeness, it should be said that when the dynamics is a stochastic differential equation, the committor function $q(\mathbf{x})$ is the solution of the Dirichlet problem~\citep{E2005transition,thiede2019galerkin}.

To illustrate the concept of predictability margin introduced in Sec.~\ref{sec:intro}, we choose the value $T$=200 in non-dimensional time units, which is slightly larger than the period of the limit cycle (the \enquote{natural} periodicity of strong El-Ni\~{n}o events which is 186), and of the order of the Lyapunov time.
This choice guarantees that for the deterministic dynamics, $\sigma=0$, each trajectory starting in one point of the limit cycle almost certainly will reach the threshold $x_c=-1$. \textcolor{black}{Note that for $\sigma=0$ the same behavior would occur for any $T$ larger than the limit cycle period. For $T$ smaller than this period, there would be a portion of the cycle that would not reach the threshold in the given time. Instead, for $\sigma\neq 0$, changing $T$ has the same effect on the results as changing $\sigma$ as it allows the noise more or less time to cause a transition.}

\subsection{Description of the committor function: deterministic and probabilistic predictability}\label{sec:committorcuts}

\begin{figure*}[ht]
\centering
\subfloat[][\emph{$\sigma=0$} \label{DeterministicX111}]
	{\includegraphics[width=.30\textwidth]{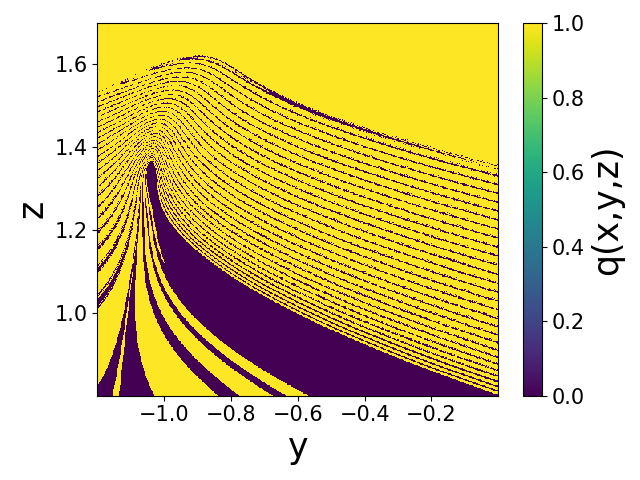} }\quad
\subfloat[][\emph{$\sigma=0.00005$}\label{X111Sigma00005}]
	{\includegraphics[width=.30\textwidth]{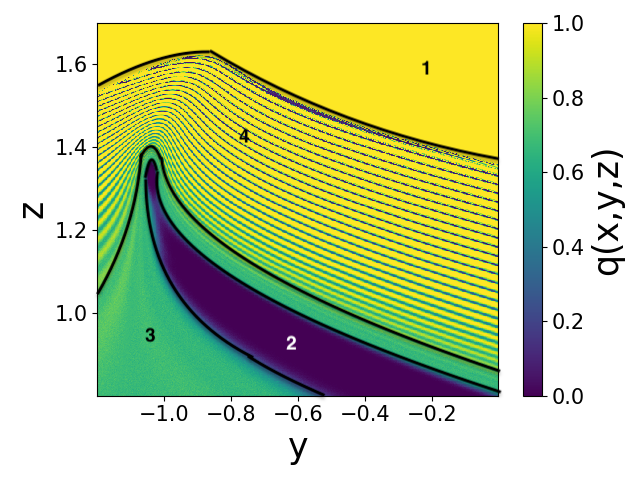}}\quad
\subfloat[][$\sigma=0.001$\label{X111Sigma001}]
	{\includegraphics[width=.30\textwidth]{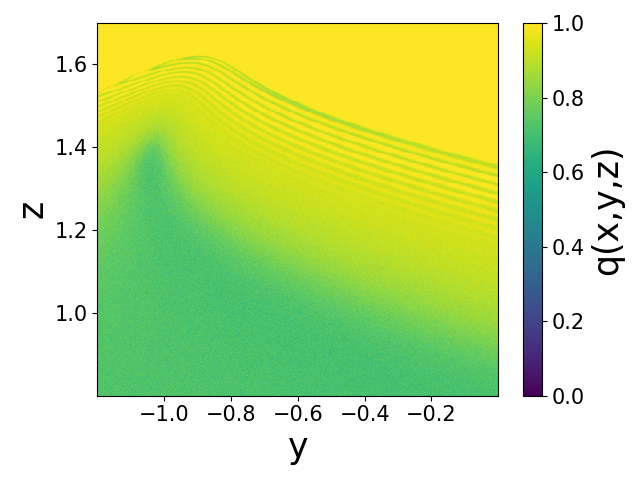} }
\caption{Color plot of the committor function $q(x, y, z)$ in the plane $x=-2.8310$, for three values of the noise amplitude, $\sigma = 0$ (left, deterministic), $0.00005$ (middle) and $0.001$ (right). Regions with uniform $q=0$ or 1 values correspond to \emph{deterministic predictability}, smooth regions with $0 < q < 1$ to \emph{probabilistic predictability}, and regions with sensitive dependence on initial conditions to unpredictable parts of phase space.}
\label{X111}
\end{figure*}
Figure~\ref{X111} shows the committor function $q(\mathbf{x})$, for different values of $\sigma$.
As $q$ is a function of 3 variables  $(x,y,z)$, we have chosen to represent cuts of $q$ in different planes.
We will discuss in detail the cut of $q$ along the plane $x=-2.831$ (Fig.~\ref{X111}) and also cuts along the planes $y=-1.1580$ and $z=1.3409$ (Fig.~\ref{Y} and Fig.~\ref{Z300}, respectively).
To compute the committor function $q(\mathbf{x})$ on the different planes we adopted the following strategy:
\begin{enumerate}
\item Discretize the plane into $K=L\times L$ cells $\mathcal{C}_k$.
\item For each cell $\mathcal{C}_k$, generate $N=1000$ trajectories starting from a point $\mathbf{x} \in \mathcal{C}_k$.
\item Count the number of trajectories $N_1$ that reach the threshold $x_c$ before time $T$.
\item Estimate $q(\mathbf{x})$ for $\mathbf{x}\in \mathcal{C}_k$ as $\hat{q}(\mathbf{x})=\frac{N_1}{N}$.
\end{enumerate}
This method is less efficient, from a computational point of view, than the one based on Eq.~\eqref{eq:Committor_Trajectory}.
In fact, in the former we only use the information carried by the initial condition while in the latter we use the information carried by a much more significant part of the trajectory.
The committor function computed from long trajectories using Eq.~\eqref{eq:Committor_Trajectory} will be discussed in Sec.~\ref{sec:committorenso}~\ref{sec:committor3d}.

\subsubsection{Committor Function for the deterministic dynamics ($\sigma=0$)} \label{sec:committor_deterministic}
In the deterministic case ($\sigma=0$, Fig.~\ref{DeterministicX111}), as the future is completely determined by the initial condition, $q$ can only take values $0$ or $1$.
On Fig.~\ref{DeterministicX111}, we can distinguish three regions corresponding to two very different situations.
First, two regions correspond to uniform values of the committor function:
in the yellow area $q=1$, when trajectories reach the threshold within a time $T$, corresponding to large values of $z$; in a thick purple band $q=0$, when no trajectory reaches the threshold.
In those two regions, the occurrence of strong El-Ni\~{n}o events is easily predicted.
Everywhere else, we see very fine filaments of alternating yellow and purple values.
In this third region, because of the sensitive dependence on the initial conditions, a \textbf{small but finite} initial perturbation, of the order of $1\%$ of the values of $x$ or $y$, leads to a different outcome.
In this region, the occurrence of strong El-Ni\~{n}o events is very difficult to predict.

We stress that a precise definition of this third area is not intrinsic, it depends on the actual precision with which the values of $x$ and $y$ can be measured. However the distinction between areas with easy predictability and areas with difficult predictability, might be crucial at a practical level. \textcolor{black}{This important practical point will be discussed more in section  \ref{sec:committorenso}.\ref{predictability_regions_scales}.}

One might ask what relationship exists between the regions outlined above and the basins of attraction of the system.
However, this relationship is less trivial than one might expect.
Although some regions reflect the structure of the basins of attraction, this is not true in general.
In fact, there are points in the basin of attraction of the strange attractor which pass the threshold before reaching the strange attractor, as well as points in the basin of attraction of the limit cycle which do not reach the threshold within the time T.

\subsubsection{Committor Function for $\sigma\neq0$} \label{sec:committor_stochastic}
Figures~\ref{X111Sigma00005} and~\ref{X111Sigma001} show the committor function in the case where a finite noise amplitude $\sigma\neq 0$ is considered.
As can be seen by comparing Figs.~\ref{DeterministicX111},~\ref{X111Sigma00005} and~\ref{X111Sigma001}, adding a small noise blurs the visible structures of the deterministic case.
For larger noise values ($\sigma=10^{-3}$), Fig.~\ref{X111Sigma001} shows that the committor function looks smooth nearly everywhere (mathematically it is smooth everywhere, smooth here is used qualitatively and means with mild variations).
This means that the \emph{deterministic predictability is lost} for most initial conditions as ($0 < q < 1$).
Then one cannot expect to predict the outcome in the way of a deterministic forecast.
However, the occurrence of strong El-Ni\~{n}o events is \emph{probabilistically predictable}: the value of the probability can be determined in practice with an excellent precision as it changes very slowly when one changes the initial conditions.
It can also be seen on the figure that the occurrence of strong El-nino events is frequent ($q>0.6$ almost everywhere).
This is an indication that for such a value of $\sigma$ we are in the noise-dominated regime.

The most interesting case is probably the one with the intermediate noise amplitude $\sigma = 0.00005$.
In Fig.~\ref{X111Sigma00005}, we delineate 4 regions: first, two regions of perfect (deterministic) predictability, where the event will occur with probability very close to $1$ (region 1) or to $0$ (region 2).
Second, there exists a \textit{probabilistically predictable region} (region 3) with good (probabilistic) predictability properties, where a value $0<q<1$ can clearly be predicted with very mild dependence with respect to the initial conditions.
Finally, the region 4, which is unpredictable in practice.
In this region, the strong dependence with respect to the initial condition prevents any practical prediction, either deterministic or probabilistic, of the precise value of $q$.
While regions 1, 2 and 4 are reminiscent of their deterministic counterparts (Fig.~\ref{DeterministicX111}), region 3 is not.
Instead, the behavior in this region is similar to the strong noise case shown in Fig.~\ref{X111Sigma001}.
It is a region where the stochasticity is large enough to smooth out the deterministic values of $q$.
The fact that it occurs even at very low noise amplitude is probably related to extremely unstable parts of the phase space, for instance for trajectories passing close to unstable fixed points or orbits. \textcolor{black}{We study the dynamical mechanisms which lead to this area with good probabilistic predictability in section \ref{sec:committorenso}.\ref{sec:lyapunov}.}

\textcolor{black}{We stress that the existence of these 4 types of predictability regions, and especially the new and most interesting \textit{probabilistically predictable region} (region 3), should be generic for most prediction problems in climate dynamics. In the present section, we have assumed a perfect model and implicitly assumed an observation scale for determining the initial conditions. In section  \ref{sec:committorenso}.\ref{predictability_regions_scales}, we will discuss the effect of a changing observation scale, and of possible model errors, on the meaning of these 4 types of predictability regions.}

\begin{figure*}[ht]
\centering
\subfloat[][\emph{$\sigma=0.00005$}\label{YSigma00005}]
	{\includegraphics[scale=0.4]{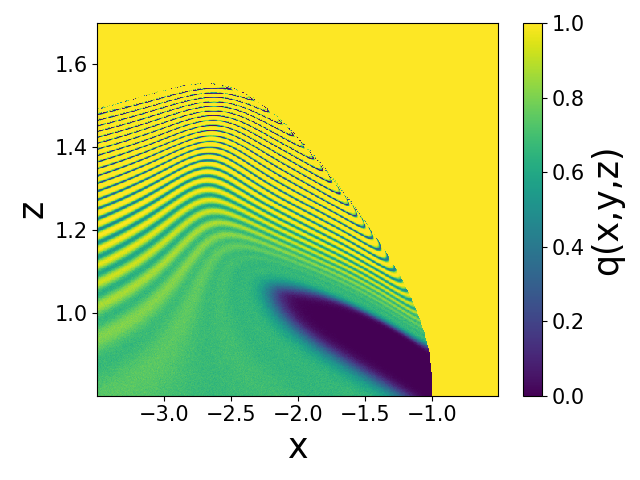}}\quad
\subfloat[][$\sigma=0.001$\label{YSigma001}]
	{\includegraphics[scale=0.4]{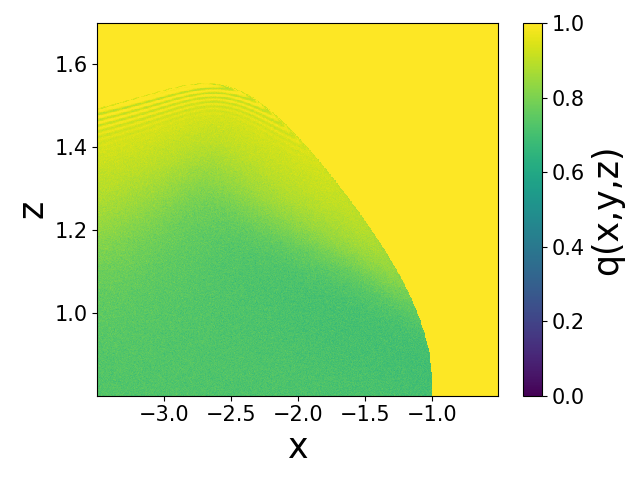} }
\caption{Color plot of the committor function $q(x, y, z)$ in the plane $y=-1.1580$, for $\sigma = 0.00005$ (left) and $0.001$ (right).}\label{Y}
\end{figure*}
\begin{figure*}[ht]
\centering
\subfloat[][\emph{$\sigma=0.00005$}\label{Z300Sigma00005}]
	{\includegraphics[scale=0.4]{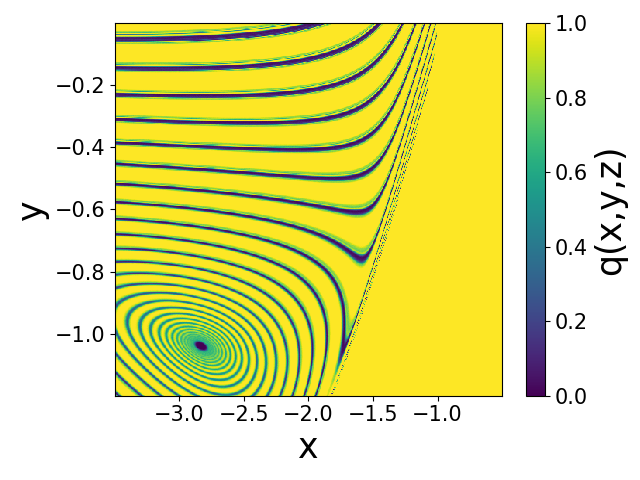}}\quad
\subfloat[][$\sigma=0.001$\label{Z300Sigma001}]
	{\includegraphics[scale=0.4]{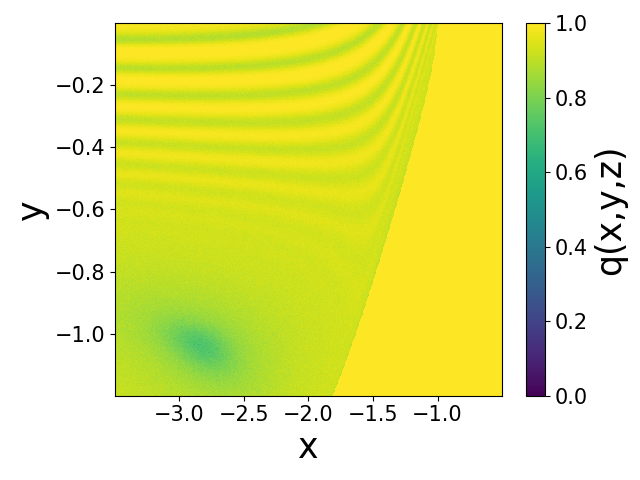} }
\caption{Color plot of the committor function $q(x, y, z)$ in the plane $z=1.3409$, for $\sigma =0.00005$ (left) $0.001$ (right).}
\label{Z300}
\end{figure*}

\subsection{Dynamical characterization of the probabilistically predictable region}\label{sec:lyapunov}

In order to understand the reason for which the probabilistically predictable region arises, it can be useful to introduce a quantity which characterizes sensitivity to small perturbations.
This quantity is the largest finite time Lyapunov exponent and it is a measure of the sensitivity to initial conditions for the deterministic system.
Let us consider two trajectories of the deterministic system ($\sigma=0$) with initial conditions $\mathbf{x}$ and $\mathbf{x+\delta x}$ and let $\Delta(t)$ be the value of the euclidean distance between the two trajectories at time $t$.
The largest Lyapunov exponent $\Lambda_L$ is defined as
\begin{equation}
  \Lambda_L=\lim\limits_{t\to+\infty}\lim\limits_{\Delta(0)\to0}\frac{1}{t}\log{\left(\frac{\Delta(t)}{\Delta(0)}\right)}.
\end{equation}
Since we are dealing with predictions with a time horizon $T$, we believe it is more appropriate to define a finite-time version of $\Lambda_L$.
Hence, we compute the largest finite time Lyapunov exponent $\lambda_L$ as $\lambda_L=\frac{1}{T}\log{(\frac{\Delta(T)}{\Delta(0)})}$.
Note that the initial perturbations have to be considered small but finite as we have taken $\mathbf{x}$ and $\mathbf{x + \delta x}$ into the same cell $\mathcal{C}_k$.
Positive values of $\lambda_L$ mean that the distance between the trajectories grows exponentially.
This quantity is shown for the Jin-Timmermann model in the $x=-2.831$ plane in Fig.~\ref{FiniteLyapunovTime}.
\begin{figure}[ht]
\centering
\includegraphics[scale=0.4]{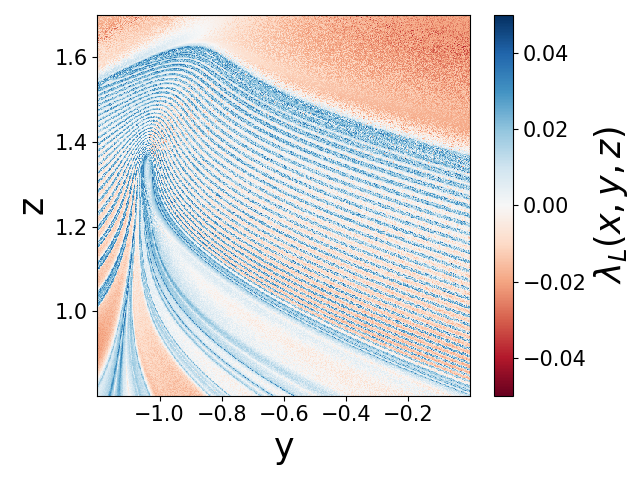}
\caption{Color plot of the maximum finite time Lyapunov exponent $\lambda_L$ as a function of the initial condition $(x, y, z)$ in the plane $x=-2.831$.}
\label{FiniteLyapunovTime}
\end{figure}
Comparing this figure with Fig.~\ref{DeterministicX111} is enlightening: the regions of perfect predictability (1 and 2) are associated to values of $\lambda_L$ which are either negative or close to $0$.
Moreover, $\lambda_L < 0$ for almost all regions for which the deterministic committor is equal to $1$ (see the two thick yellow bands at the bottom left of Fig.\ref{DeterministicX111}).
From this description, it seems that $\lambda_L$ is not the correct quantity to explain the emergence of the probabilistically predictable region.
However, from a careful analysis of Fig.~\ref{FiniteLyapunovTime} it can be noted that there is a region, close to the left boundary of region 2, for which the values of $\lambda_L$ are positive also for points $\mathbf{x}$ such that $q(\mathbf{x})=1$.
It means that this region is quite unstable with respect to small initial perturbations.
Since this region belongs to the \emph{probabilistically predictable region}, it is reasonable to say that the appearance of region 3 is related with this instability.
In fact, although the instability region is a subset of the \emph{probabilistically predictable region}, it should be noted that for $\sigma\neq0$ the system is perturbed at any time.
The ensemble of these small perturbations gives rise to a finite perturbation which could explain the growth of the region of instability.

To reinforce this conjecture, we compute the averaged value of the euclidean distance between two different trajectories $\mathbf{x}_1(t)$ and $\mathbf{x}_2(t)$, with the same initial conditions but different realisations of the noise:$\langle d_{\text{max}} \rangle = \langle \max_{t\in[0,T]} \|\mathbf{x}_1(t)-\mathbf{x}_2(t)\|^2 \rangle$.
\begin{figure}[ht]
\centering
\includegraphics[scale=0.4]{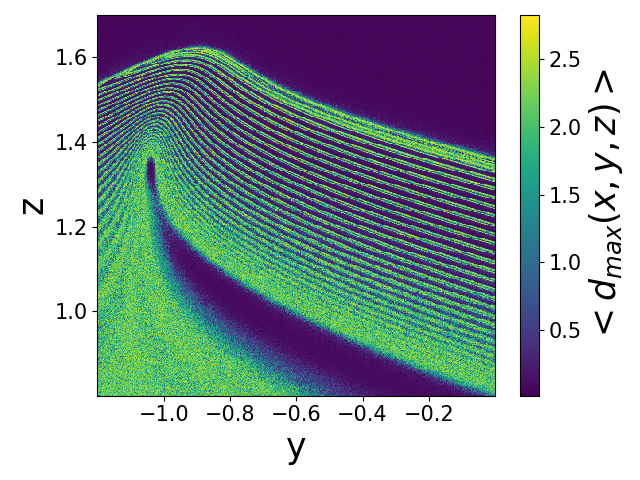}
\caption{Color plot of the average value of the maximum distance $\langle d_{\text{max}} \rangle$ as a function of the initial condition $(x, y, z)$ in the plane $x=-2.831$, computed for $\sigma=0.00005$.}
\label{distmax}
\end{figure}
Figure~\ref{distmax} shows $\langle d_{\text{max}} \rangle$ as a function of the initial condition in the plane $x=-2.831$ for $\sigma=0.00005$: it can be seen that the typical values of $d_{\text{max}}$ are larger in the probabilistically predictable region than in other regions.
This means that trajectories starting in this region are more sensitive to noise induced perturbations than trajectories starting in different regions of phase space.
Therefore, the emergence of the \emph{probabilistically predictable region} is qualitatively associated to an instability present in the deterministic system which is accentuated as the amplitude of the perturbations increases.

\subsection{Committor function computed from long trajectories}\label{sec:committor3d}
In this section we discuss the committor function computed from an ensemble of long trajectories by means of Eq.~\eqref{eq:Committor_Trajectory}.
The motivation is that the very precise strategy adopted in Sec.~\ref{sec:committorenso}\ref{sec:committorcuts} is unlikely to be adapted for real-world problems as well, because it requires a dynamical model and has a very large computational cost.
Indeed, it requires to sample an ensemble of trajectories for every point in the phase space.
For a data-based approach, it is usually possible to observe only the evolution of a trajectory (or of an ensemble of trajectories) over a very long time.
However, individual trajectories, regardless of their length, do not fill the whole phase space.
Indeed, they usually concentrate on the region where the invariant measure of the stochastic system is concentrated.
Hence, the strategy adopted in Sec.~\ref{sec:committorenso}\ref{sec:committorcuts} is appropriate for computing the committor function in an arbitrary plane while the use of Eq.~\eqref{eq:Committor_Trajectory} allows the computation of the committor function in the region where the invariant measure is concentrated.

The committor function computed using Eq.~\eqref{eq:Committor_Trajectory} is shown in Fig.~\ref{Committor3D2}, for the same three values of the noise amplitude as in Sec.~\ref{sec:committorenso}\ref{sec:committorcuts}.
\begin{figure*}[ht]
\centering
\subfloat[][\emph{$\sigma=0$} \label{DeterministicCommittor3D}]
	{\includegraphics[width=.3\textwidth]{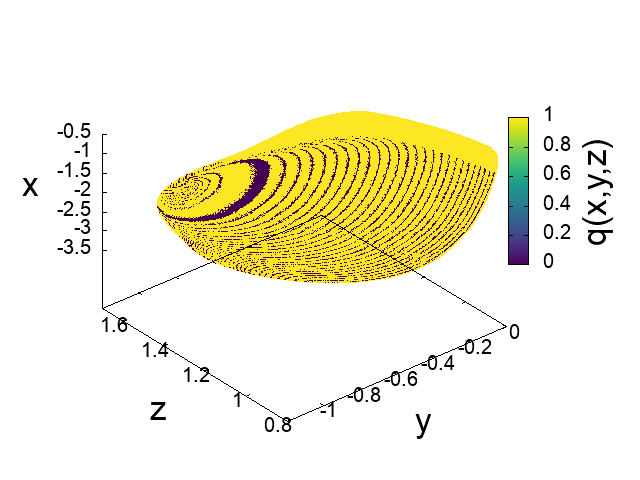} }\quad
\subfloat[][\emph{$\sigma=0.00005$}\label{Committor3D}]
	{\includegraphics[width=.3\textwidth]{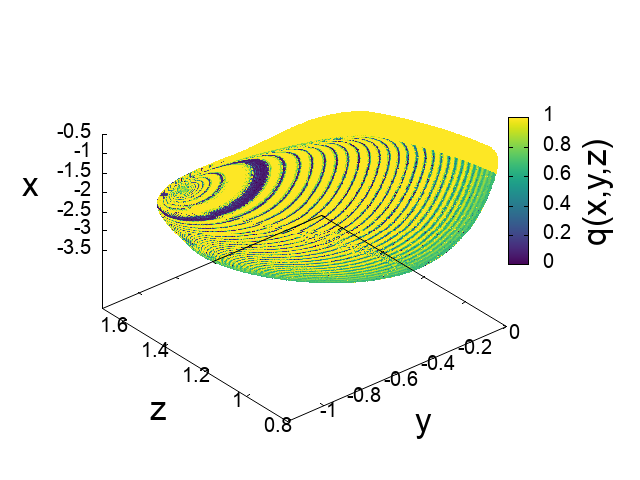}}\quad
\subfloat[][\emph{$\sigma=0.001$}\label{Committor3DSigma001}]
	{\includegraphics[width=.3\textwidth]{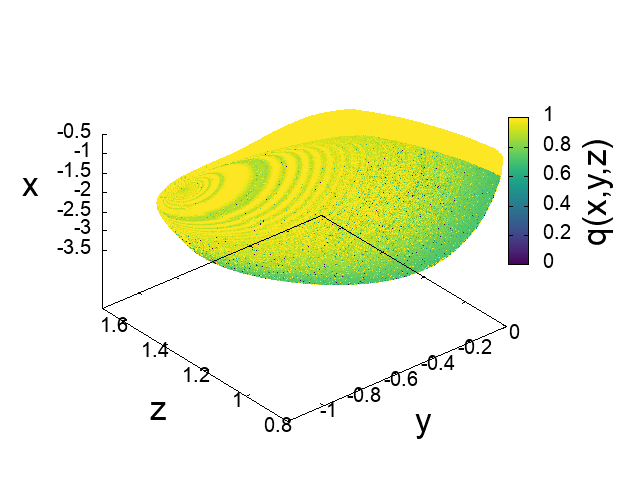}}
\caption{Color map of the committor function $q(x,y,z)$ in correspondence to the most visited region of phase space, for $\sigma = 0$ (left, deterministic), $0.00005$ (middle) and $0.001$ (right).}
\label{Committor3D2}
\end{figure*}
For this figure, we used 10 000 trajectories of length $\mathcal{T}=10^4$ (nondimensional time units) initialized in the strange attractor.
Note that $\mathcal{T}$ is $10$ times bigger than the first exit time $E[\tau_c]$ for $\sigma=0.00005$ and $100$ times greater than $E[\tau_c]$  for $\sigma = 0.001$ (see Fig.~\ref{MeanExitTime}).
This guarantees that the trajectories will be distributed according to the invariant measure of the system.

As already mentioned, Fig.~\ref{Committor3D2} shows that the trajectories do not cover all the phase space but they are concentrated in a certain region.
Furthermore, we can see that it is more appropriate to call it a manifold rather than a region.
In fact, if it were an object of dimension $3$, its intersection with a plane should define an area on that plane.
Instead, it appears that the intersections between the object and planes are lines rather than areas.
This is illustrated in Fig.~\ref{fig:Intersection} which shows an intersection between the object in Fig.~\ref{Committor3D} and the plane $x=-2.831$ (the same plane as in Fig.~\ref{X111}).
This leads us to conclude that the trajectories are distributed over a manifold of dimension smaller than $3$.
\begin{figure}[ht]
\centering
\includegraphics[scale=0.4]{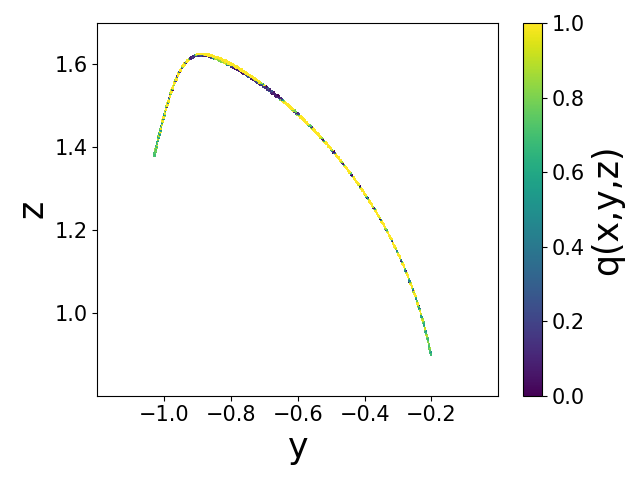}
\caption{Color plot of the committor function for $\sigma=0.00005$ at the intersection of the manifold with the plane $x=-2.831$.}
\label{fig:Intersection}
\end{figure}

The comparison between Fig.~\ref{fig:Intersection} and Fig.~\ref{X111Sigma00005} allows us to make another important remark on the committor function on the manifold.
As can be seen in Fig.~\ref{fig:Intersection}, the committor function takes values between $0$ and $1$ at the two ends of the line.
For the rest, the line is made up of segments on which $q$ takes the values $0$ or $1$.
It is straightforward to recognize that the two ends of the line belong to the probabilistically predictable region while the central part of the line belongs to region 4.
This highlights that the committor function computed from long trajectories provides useful information for many of the states of the system, that is, for all typical states.
However, for atypical conditions where we have little or no information, using $q$ to make predictions can lead to erroneous results.

Having made these necessary considerations, we can continue the description of the committor computed from long trajectories.
By comparing Fig.~\ref{Committor3D2} with Fig.~\ref{TwoAttractors} we can immediately identify the strange attractor in the manifold on which the committor function is represented.
The limit cycle is more difficult to visualize but its presence can be deduced from the spiral behavior present in the region inside the strange attractor and from the shape of the manifold boundaries that follow the shape of the limit cycle in Fig.~\ref{TwoAttractors}.

The qualitative structure of the committor function on the manifold is similar to the one observed on plane cuts: in the deterministic case, we observe regions of perfect predictability, and regions where the sensitivity to initial conditions make it unpredictable in practice.
When the noise is sufficiently strong, regions of probabilistic predictability appear, where a finite value of the probability of a strong El Niño event $0 < q < 1$ depends only mildly on the initial conditions (Fig.~\ref{Committor3DSigma001}).
It can be noted, however, that the intermediate case, analogous to Fig.~\ref{X111Sigma00005}, with coexistence of a region of deterministic predictability, a region of probabilistic predictability, and an unpredictable region, is more difficult to observe in this visualization.

Finally, we underline that the region of unpredictability, made by thin filaments where the committor is a highly fluctuating function, emphasizes again how the two attractors are intertwined in a complex way.

\subsection{The effect of the observation scale and of the model resolution on the predictability areas}\label{predictability_regions_scales}

\textcolor{black}{In section section \ref{sec:committorenso}.\ref{sec:committorcuts}.1 and in section \ref{sec:committorenso}.\ref{sec:committorcuts}.2, we have described 4 regions depending on their predictability properties. In Fig.~\ref{X111Sigma00005}, the distinction between the \textit{probabilistically predictable region} (region 3) with good (probabilistic) predictability properties, and region 4 which is unpredictable in practice, is very clear visually. However we stress again that this distinction, of high practical relevance, is not intrinsic and that the probabilistic predictability properties can be characteristic of a system or emerge due to ignorance of the initial conditions or the lack of a sufficiently precise model}. 

\textcolor{black}{Indeed, for practical applications, the initial condition is known only up to some errors. Let us for instance assume that the initial condition is known up to a given error that is characterized by a scale $l_o$. For this case, the practitioner would see an effective committor which would be the real committor function (for instance as represented in Fig.~\ref{X111Sigma00005})) which would be spatially averaged at the observation scale $l_o$. If $l_o$ is larger than the typical size of the filaments in region 4, then the unpredictable region will appear practically as having a smooth committor function at the scale  $l_o$, and the practitioner will not see qualitative differences between the regions 3 and 4. However, if the practitioner is able to improve her knowledge of the initial condition, and decrease $l_o$ below the size of the filaments, she will learn that the value of the committor function varies abruptly with the initial condition in the unpredictable region 4, while it will maintain smooth and more or less constant values when changing $l_o$ in the \textit{probabilistically predictable region} (region 3). The prediction in the \textit{probabilistically predictable region} remains robust, although probabilistic.}
 
\textcolor{black}{In the case of an imperfect model, for instance due to a finite resolution $l_r$, the amplitude of the effective noise should decrease when $l_r$ decrease. This idea of somehow too simplistic, as the deterministic term itself should be slightly modified. However it helps understand qualitatively the effect of the model resolution on the committor function. If the practitioner is able to improve his model resolution by decreasing $l_r$, he will be able to progressively decrease the noise amplitude and uncover the fine structure of the committor function in the unpredictable region 4, and he will observe abrupt variations of the committor function when the initial condition is changed. By contrast, in the region of \textit{probabilistically predictable region}, the change of the resolution will not lead to qualitative or important quantitative changes for the committor function, the prediction will be robust even with coarser resolution.}

\section{Conclusion}

In this paper, we have introduced a mathematical concept, the \emph{committor function}, encoding the probability that an event occurs within a given time, conditioned on the state of the system at the time of prediction.
We believe it is an appropriate concept for many prediction problems in climate science in a range of time scales which we call the \emph{predictability margin}.
It corresponds to timescales for which a deterministic description of the system is no longer relevant, because of the sensitive dependence to initial conditions, but for which more precise probabilistic predictions than the climatological one can be made, because the system has not yet forgotten completely the initial condition.

In the context of a simple, low-dimensional stochastic model, the Jin-Timmermann model, in a regime of coexistence of a limit cycle and a strange attractor found by~\cite{guckenheimer2017predictability}, we have shown that noise could induce transitions between the two attractors.
These transitions correspond to regime shifts regarding the occurrence of strong El Niño events, which are periodic in the limit cycle, with a return time close to 15 years, and which do not occur at all in the strange attractor (in the deterministic case).
In the stochastic case, the occurrence of strong El Niño events therefore becomes random, and the waiting times follow a Poisson statistics.

In this example, we have shown that the probability of occurrence of strong El Niño events had different predictability properties depending on the state of the system at the time of prediction.
The most important result is that there exist regions of \emph{probabilistic predictability}, where the event has a finite probability of occurring $0 < q < 1$, and this probability does not depend sensitively on the initial state, and regions of \emph{probabilistic unpredictability} where the probability changes a lot if one changes by a small and finite amount the initial condition.
We expect the existence of this dichotomy between probabilistically predictable and probabilistically unpredictable regions to be a generic feature for climate prediction problems at the predictability margin.
We stress that this notion depends on the precision with which the initial condition can be assessed and on the model resolution. \textcolor{black}{We also discussed that in the regions of \emph{probabilistic predictability}, the prediction is expected to be more robust that in the regions of \emph{probabilistic unpredictability}, when either the initial conditions, the observation scale, or the model resolution scale, are changed.}  

We have also discussed the methodological aspects for computing the committor functions.
For our example, a small stochastic perturbation of a chaotic deterministic system, we have computed the committor function using two approaches.
First, by direct sampling of ensembles of initial conditions close to any point in phase space, and second, through a data based approach using observed trajectories.
As soon as the number of degrees of freedom increases, the first method will become impossible to use in practice, because of the numerical cost.
The second method may sometimes be associated with sampling issues, as one can get meaningful results only for the parts of the phase space that have been visited many times.
Another method mentioned in Sec.~\ref{sec:committorenso}, would be to solve a backward Kolmogorov equation by employing classical finite difference methods.
This method is impractical for systems with more than a few degrees of freedom.
To be able to sample efficiently committor functions in large dimensions, more efficient data-based methods will be necessary, relying either on classical statistical methods or machine learning methods~\citep{Lucente2019}. Furthermore, such data-based methods are also effective for solving backward Kolmogorov equations~\citep{thiede2019galerkin,finkel2021learning,lucente2021coupling}.

\textcolor{black}{A natural question is to understand the difficulty to compute committor functions using the best up to date atmosphere or climate models. In a series of forthcoming papers, we develop a method based on machine learning for computing committor functions for the prediction of mid-latitude extreme heat waves using data from one of the best current climate model, CESM2.1, the model used by NCAR for the CMIP experiments.}

\begin{acknowledgments}
  The work of D.~Lucente was funded through the ACADEMICS grant of IDEXLYON, project of the Université de Lyon, PIA operated by ANR-16-IDEX-0005. This project has received funding from the European Union's Horizon 2020 research and innovation programme under the Marie Sk\l odowska-Curie grant agreement No 753021 and the ANR grant SAMPRACE, project ANR-20-CE01-0008-01. 
  Computer time was provided by the \enquote{P\^ole Scientifique de Mod\'elisation Num\'erique} in Lyon.
  We acknowledge useful scientific discussions with P.~Abry, P.~Borgnat, C.-E.~Bréhier, J.~Finkel, T. Lelièvre, J.~Rolland and J.~Weare.
\end{acknowledgments}

\datastatement
This theoretical work does not use external datasets. The datasets on which this paper is based as well as the numerical code are not archived on a public data repository. All the information for reproducing the results are provided in the main body. Additionally, the corresponding author may provide the data upon reasonable request.

\bibliographystyle{ametsoc2014}
\bibliography{Bibliography}

\begin{thebibliography}{55}
\providecommand{\natexlab}[1]{#1}
\providecommand{\url}[1]{\texttt{#1}}
\renewcommand{\UrlFont}{\rmfamily}
\providecommand{\urlprefix}{URL }
\expandafter\ifx\csname urlstyle\endcsname\relax
  \providecommand{\doi}[1]{doi:\discretionary{}{}{}#1}\else
  \providecommand{\doi}{doi:\discretionary{}{}{}\begingroup
  \urlstyle{rm}\Url}\fi
\providecommand{\eprint}[2][]{\url{#2}}

\bibitem[{AghaKouchak et~al.(2012)AghaKouchak, Easterling, Hsu, Schubert,, and
  Sorooshian}]{aghakouchak2012extremes}
AghaKouchak, A., D.~Easterling, K.~Hsu, S.~Schubert, and S.~Sorooshian, 2012:
  \textit{Extremes in a changing climate: detection, analysis and uncertainty},
  Vol.~65. Springer Science \& Business Media.

\bibitem[{Bauer et~al.(2015)Bauer, Thorpe,, and Brunet}]{Bauer2015}
Bauer, P., A.~Thorpe, and G.~Brunet, 2015: The quiet revolution of numerical
  weather prediction. \textit{Nature}, \textbf{525~(7567)}, 47,
  \doi{10.1038/nature14956}.

\bibitem[{Bouchet and Reygner({2016})Bouchet, and
  Reygner}]{Bouchet_Reygner_2016}
Bouchet, F., and J.~Reygner, {2016}: {Generalisation of the Eyring-Kramers
  Transition Rate Formula to Irreversible Diffusion Processes}.
  \textit{{ANNALES HENRI POINCARE}}, \textbf{{17}~({12})}, {3499--3532},
  \doi{{10.1007/s00023-016-0507-4}}.

\bibitem[{Bouchet et~al.(2019)Bouchet, Rolland,, and Simonnet}]{Bouchet2019}
Bouchet, F., J.~Rolland, and E.~Simonnet, 2019: A rare event algorithm links
  transitions in turbulent flows with activated nucleations. \textit{Phys. Rev.
  Lett.}, \textbf{122}, 074\,502, \doi{10.1103/PhysRevLett.122.074502}.

\bibitem[{Castiglione et~al.(2008)Castiglione, Falcioni, Lesne,, and
  Vulpiani}]{castiglione2008chaos}
Castiglione, P., M.~Falcioni, A.~Lesne, and A.~Vulpiani, 2008: \textit{Chaos
  and coarse graining in statistical mechanics}. Cambridge University Press
  Cambridge.

\bibitem[{Chekroun et~al.(2011)Chekroun, Kondrashov,, and Ghil}]{Chekroun2011a}
Chekroun, M.~D., D.~Kondrashov, and M.~Ghil, 2011: {Predicting stochastic
  systems by noise sampling, and application to the El Ni{\~n}o-Southern
  Oscillation}. \textit{Proc. Natl. Acad. Sci. U.S.A.}, \textbf{108~(29)},
  11\,766, \doi{10.1073/pnas.1015753108}.

\bibitem[{Clarke(2008)}]{clarke2008introduction}
Clarke, A.~J., 2008: \textit{An introduction to the dynamics of El Ni{\~n}o \&
  the Southern Oscillation}. Academic Press.

\bibitem[{Cobb et~al.(2003)Cobb, Charles, Cheng,, and Edwards}]{Cobb2003}
Cobb, K.~M., C.~D. Charles, H.~Cheng, and R.~L. Edwards, 2003: {El
  Ni{\~n}o/Southern Oscillation and tropical Pacific climate during the last
  millennium}. \textit{Nature}, \textbf{424~(6946)}, 271--276,
  \doi{10.1038/nature01779}.

\bibitem[{Cobb et~al.(2013)Cobb, Westphal, Sayani, Watson, Di~Lorenzo, Cheng,
  Edwards,, and Charles}]{Cobb2013}
Cobb, K.~M., N.~Westphal, H.~R. Sayani, J.~T. Watson, E.~Di~Lorenzo, H.~Cheng,
  R.~Edwards, and C.~D. Charles, 2013: {Highly variable El Ni{\~n}o--Southern
  Oscillation throughout the Holocene}. \textit{Science}, \textbf{339~(6115)},
  67--70, \doi{10.1126/science.1228246}.

\bibitem[{Coumou and Rahmstorf(2012)Coumou, and Rahmstorf}]{coumou2012decade}
Coumou, D., and S.~Rahmstorf, 2012: A decade of weather extremes. \textit{Nat.
  Clim. Change}, \textbf{2~(7)}, 491, \doi{10.1038/nclimate1452}.

\bibitem[{Dijkstra(2013)}]{dijkstra2013nonlinear}
Dijkstra, H.~A., 2013: \textit{Nonlinear climate dynamics}. Cambridge
  University Press.

\bibitem[{Dykman and Ryvkine(2005)Dykman, and Ryvkine}]{dykman2005activated}
Dykman, M.~I., and D.~Ryvkine, 2005: Activated escape of periodically modulated
  systems. \textit{Phys. Rev. Lett.}, \textbf{94~(7)}, 070\,602,
  \doi{10.1103/PhysRevLett.94.070602}.

\bibitem[{E et~al.(2005)E, Ren,, and Vanden-Eijnden}]{E2005transition}
E, W., W.~Ren, and E.~Vanden-Eijnden, 2005: Transition pathways in complex
  systems: Reaction coordinates, isocommittor surfaces, and transition tubes.
  \textit{Chemical Physics Letters}, \textbf{413~(1-3)}, 242--247,
  \doi{10.1016/j.cplett.2005.07.084}.

\bibitem[{Fedorov et~al.(2006)Fedorov, Dekens, McCarthy, Ravelo, DeMenocal,
  Barreiro, Pacanowski,, and Philander}]{Fedorov2006}
Fedorov, A., P.~Dekens, M.~McCarthy, A.~Ravelo, P.~DeMenocal, M.~Barreiro,
  R.~Pacanowski, and S.~Philander, 2006: {The Pliocene paradox (mechanisms for
  a permanent El Ni{\~n}o)}. \textit{Science}, \textbf{312~(5779)}, 1485--1489,
  \doi{10.1126/science.1122666}.

\bibitem[{Feng and Dijkstra(2017)Feng, and Dijkstra}]{feng2017climate}
Feng, Q.~Y., and H.~A. Dijkstra, 2017: {Climate network stability measures of
  El Ni{\~n}o variability}. \textit{Chaos}, \textbf{27~(3)}, 035\,801,
  \doi{10.1063/1.4971784}.

\bibitem[{Field et~al.(2012)Field, Barros, Stocker,, and
  Dahe}]{field2012managing}
Field, C.~B., V.~Barros, T.~F. Stocker, and Q.~Dahe, 2012: \textit{Managing the
  risks of extreme events and disasters to advance climate change adaptation:
  special report of the intergovernmental panel on climate change}. Cambridge
  University Press.

\bibitem[{Finkel et~al.(2020)Finkel, Abbot,, and Weare}]{finkel2020path}
Finkel, J., D.~S. Abbot, and J.~Weare, 2020: Path properties of atmospheric
  transitions: illustration with a low-order sudden stratospheric warming
  model. \textit{Journal of the Atmospheric Sciences}, \textbf{77~(7)},
  2327--2347.

\bibitem[{Finkel et~al.(2021)Finkel, Webber, Gerber, Abbot,, and
  Weare}]{finkel2021learning}
Finkel, J., R.~J. Webber, E.~P. Gerber, D.~S. Abbot, and J.~Weare, 2021:
  Learning forecasts of rare stratospheric transitions from short simulations.
  \textit{Monthly Weather Review}, \textbf{149~(11)}, 3647--3669.

\bibitem[{Freidlin and Wentzell(2012)Freidlin, and
  Wentzell}]{Freidlin_Wentzell}
Freidlin, M.~I., and A.~D. Wentzell, 2012: \textit{{Random Perturbations of
  Dynamical Systems}}. Springer-Verlag, 3dr ed. New York.

\bibitem[{Gardiner et~al.(1985)}]{gardiner1985handbook}
Gardiner, C.~W., and Coauthors, 1985: \textit{Handbook of stochastic methods},
  Vol.~3. Springer Berlin.

\bibitem[{Gayrard et~al.(2004)Gayrard, Bovier, Eckhoff,, and
  Klein}]{bovier2004metastability}
Gayrard, V., A.~Bovier, M.~Eckhoff, and M.~Klein, 2004: Metastability in
  reversible diffusion processes i: Sharp asymptotics for capacities and exit
  times. \textit{Journal of the European Mathematical Society}, \textbf{6~(4)},
  399--424.

\bibitem[{Graham(1987)}]{Graham1987macroscopic}
Graham, R., 1987: Macroscopic potentials, bifurcations and noise in dissipative
  systems. \textit{Fluctuations and Stochastic Phenomena in Condensed Matter},
  Springer, 1--34.

\bibitem[{Guckenheimer et~al.(2017)Guckenheimer, Timmermann, Dijkstra,, and
  Roberts}]{guckenheimer2017predictability}
Guckenheimer, J., A.~Timmermann, H.~Dijkstra, and A.~Roberts, 2017: {(Un)
  predictability of strong El Ni{\~n}o events}. \textit{Dynamics and Statistics
  of the Climate System}, \textbf{2~(1)}, dzx004, \doi{10.1093/climsys/dzx004}.

\bibitem[{Herbert and Bouchet(2017)Herbert, and Bouchet}]{Herbert2017}
Herbert, C., and F.~Bouchet, 2017: {Predictability of escape for a stochastic
  saddle-node bifurcation: when rare events are typical}. \textit{Phys. Rev.
  E}, \textbf{96}, 030\,201(R), \doi{10.1007/BF01106788}.

\bibitem[{Herring et~al.(2014)Herring, Hoerling, Peterson,, and
  Stott}]{herring2014explaining}
Herring, S.~C., M.~P. Hoerling, T.~C. Peterson, and P.~A. Stott, 2014:
  Explaining extreme events of 2013 from a climate perspective. \textit{Bull.
  Am. Meteorol. Soc.}, \textbf{95~(9)}, S1--S104,
  \doi{10.1175/1520-0477-95.9.S1.1}.

\bibitem[{Jin(1997{\natexlab{a}})}]{jin1997equatorial1}
Jin, F.-F., 1997{\natexlab{a}}: {An equatorial ocean recharge paradigm for
  ENSO. Part I: Conceptual model}. \textit{J. Atmos. Sci.}, \textbf{54~(7)},
  811--829, \doi{10.1175/1520-0469(1997)054<0811:AEORPF>2.0.CO;2}.

\bibitem[{Jin(1997{\natexlab{b}})}]{jin1997equatorial2}
Jin, F.-F., 1997{\natexlab{b}}: {An equatorial ocean recharge paradigm for
  ENSO. Part II: A stripped-down coupled model}. \textit{J. Atmos. Sci.},
  \textbf{54~(7)}, 830--847,
  \doi{10.1175/1520-0469(1997)054<0830:AEORPF>2.0.CO;2}.

\bibitem[{Kalnay(2003)}]{kalnay2003atmospheric}
Kalnay, E., 2003: \textit{Atmospheric modeling, data assimilation and
  predictability}. Cambridge University Press.

\bibitem[{Khider et~al.(2011)Khider, Stott, Emile-Geay, Thunell,, and
  Hammond}]{Khider2011}
Khider, D., L.~Stott, J.~Emile-Geay, R.~Thunell, and D.~Hammond, 2011:
  Assessing el ni{\~n}o southern oscillation variability during the past
  millennium. \textit{Paleoceanography}, \textbf{26~(3)},
  \doi{10.1029/2011PA002139}.

\bibitem[{Latif et~al.(1994)Latif, Barnett, Cane, Fl{\"u}gel, Graham,
  Von~Storch, Xu,, and Zebiak}]{latif1994review}
Latif, M., T.~P. Barnett, M.~A. Cane, M.~Fl{\"u}gel, N.~E. Graham,
  H.~Von~Storch, J.-S. Xu, and S.~E. Zebiak, 1994: {A review of ENSO prediction
  studies}. \textit{Clim. Dyn.}, \textbf{9~(4-5)}, 167--179,
  \doi{10.1007/BF00208250}.

\bibitem[{Lorenz(1969)}]{Lorenz1969a}
Lorenz, E.~N., 1969: {The predictability of a flow which possesses many scales
  of motion}. \textit{Tellus}, \textbf{21}, 289--307,
  \doi{10.1111/j.2153-3490.1969.tb00444.x}.

\bibitem[{Lucarini and B{\'o}dai(2019)Lucarini, and
  B{\'o}dai}]{lucarini2019transitions}
Lucarini, V., and T.~B{\'o}dai, 2019: Transitions across melancholia states in
  a climate model: reconciling the deterministic and stochastic points of view.
  \textit{Phys. Rev. Lett.}, \textbf{122~(15)}, 158\,701,
  \doi{10.1103/PhysRevLett.122.158701}.

\bibitem[{Lucente et~al.(2019)Lucente, Duffner, Herbert, Rolland,, and
  Bouchet}]{Lucente2019}
Lucente, D., S.~Duffner, C.~Herbert, J.~Rolland, and F.~Bouchet, 2019: {Machine
  learning of committor functions for predicting high impact climate events}.
  \textit{Proceedings of the 9th International Workshop on Climate Informatics:
  CI 2019}, J.~Brajard, A.~Charantonis, C.~Chen, and J.~Runge, Eds., NCAR,
  \doi{10.5065/y82j-f154}, \eprint{1910.11736}.

\bibitem[{Lucente et~al.(2021)Lucente, Rolland, Herbert,, and
  Bouchet}]{lucente2021coupling}
Lucente, D., J.~Rolland, C.~Herbert, and F.~Bouchet, 2021: Coupling rare event
  algorithms with data-based learned committor functions using the analogue
  markov chain. \textit{arXiv preprint arXiv:2110.05050}.

\bibitem[{Ludescher et~al.(2014)Ludescher, Gozolchiani, Bogachev, Bunde,
  Havlin,, and Schellnhuber}]{ludescher2014very}
Ludescher, J., A.~Gozolchiani, M.~I. Bogachev, A.~Bunde, S.~Havlin, and H.~J.
  Schellnhuber, 2014: {Very early warning of next El Ni{\~n}o}. \textit{Proc.
  Natl Acad. Sci. USA}, \textbf{111~(6)}, 2064--2066,
  \doi{10.1073/pnas.1323058111}.

\bibitem[{McGregor et~al.(2013)McGregor, Timmermann, England, Timm,, and
  Wittenberg}]{McGregor2013}
McGregor, S., A.~Timmermann, M.~H. England, O.~E. Timm, and A.~T. Wittenberg,
  2013: {Inferred changes in El Ni{\~n}o-Southern Oscillation variance over the
  past six centuries.} \textit{Clim. Past}, \textbf{9~(5)},
  \doi{10.5194/cp-9-2269-2013}.

\bibitem[{McPhaden et~al.(2015)McPhaden, Timmermann, Widlansky, Balmaseda,, and
  Stockdale}]{mcphaden2015curious}
McPhaden, M.~J., A.~Timmermann, M.~J. Widlansky, M.~A. Balmaseda, and T.~N.
  Stockdale, 2015: The curious case of the el ni{\~n}o that never happened: a
  perspective from 40 years of progress in climate research and forecasting.
  \textit{Bull. Amer. Meteor. Soc.}, \textbf{96~(10)}, 1647--1665,
  \doi{10.1175/BAMS-D-14-00089.1}.

\bibitem[{Metzner et~al.(2006)Metzner, Sch{\"u}tte,, and
  Vanden-Eijnden}]{metzner2006illustration}
Metzner, P., C.~Sch{\"u}tte, and E.~Vanden-Eijnden, 2006: Illustration of
  transition path theory on a collection of simple examples. \textit{The
  Journal of chemical physics}, \textbf{125~(8)}, 084\,110.

\bibitem[{Metzner et~al.(2009)Metzner, Sch{\"u}tte,, and
  Vanden-Eijnden}]{metzner2009transition}
Metzner, P., C.~Sch{\"u}tte, and E.~Vanden-Eijnden, 2009: Transition path
  theory for markov jump processes. \textit{Multiscale Modeling \& Simulation},
  \textbf{7~(3)}, 1192--1219.

\bibitem[{Miron et~al.(2021)Miron, Beron-Vera, Helfmann,, and
  Koltai}]{miron2021transition}
Miron, P., F.~Beron-Vera, L.~Helfmann, and P.~Koltai, 2021: Transition paths of
  marine debris and the stability of the garbage patches. \textit{Chaos: An
  Interdisciplinary Journal of Nonlinear Science}, \textbf{31~(3)}, 033\,101.

\bibitem[{Nooteboom et~al.(2018)Nooteboom, Feng, L{\'o}pez,
  Hern{\'a}ndez-Garc{\'\i}a,, and Dijkstra}]{nooteboom2018using}
Nooteboom, P.~D., Q.~Y. Feng, C.~L{\'o}pez, E.~Hern{\'a}ndez-Garc{\'\i}a, and
  H.~A. Dijkstra, 2018: {Using Network Theory and Machine Learning to predict
  El Ni{\~n}o}. \textit{Earth Syst. Dynam.}, \textbf{9}, 969--983,
  \doi{10.5194/esd-9-969-2018}.

\bibitem[{Novikov(1959)}]{Novikov1959}
Novikov, E., 1959: On the problem of predictability of synoptic processes.
  \textit{Izv. Acad. Sci. USSR, Geophys. Ser}, \textbf{11}, 1209--1211.

\bibitem[{Philander(1990)}]{philander1990nino}
Philander, S.~G., 1990: \textit{El Ni{\~n}o, La Ni{\~n}a, and the Southern
  Oscillation}, International geophysics series, Vol.~46. Academic Press.

\bibitem[{Ragone et~al.(2018)Ragone, Wouters,, and
  Bouchet}]{ragone2018computation}
Ragone, F., J.~Wouters, and F.~Bouchet, 2018: Computation of extreme heat waves
  in climate models using a large deviation algorithm. \textit{Proc. Natl Acad.
  Sci. USA}, \textbf{115~(1)}, 24--29, \doi{10.1073/pnas.1712645115}.

\bibitem[{Rickaby and Halloran(2005)Rickaby, and Halloran}]{Rickaby2005}
Rickaby, R. E.~M., and P.~Halloran, 2005: {Cool La Ni{\~n}a during the warmth
  of the Pliocene?} \textit{Science}, \textbf{307~(5717)}, 1948--1952,
  \doi{10.1126/science.1104666}.

\bibitem[{Roberts(2012)}]{roberts2012modify}
Roberts, A., 2012: Modify the improved euler scheme to integrate stochastic
  differential equations. \textit{arXiv preprint arXiv:1210.0933}.

\bibitem[{Roberts et~al.(2016)Roberts, Guckenheimer, Widiasih, Timmermann,, and
  Jones}]{roberts2016mixed}
Roberts, A., J.~Guckenheimer, E.~Widiasih, A.~Timmermann, and C.~K. Jones,
  2016: {Mixed-mode oscillations of El Ni{\~n}o-Southern Oscillation}.
  \textit{J. Atmos. Sci.}, \textbf{73~(4)}, 1755--1766,
  \doi{10.1175/JAS-D-15-0191.1}.

\bibitem[{Sarachik and Cane(2010)Sarachik, and Cane}]{sarachik2010nino}
Sarachik, E.~S., and M.~A. Cane, 2010: \textit{The El Nino-Southern Oscillation
  phenomenon}. Cambridge University Press.

\bibitem[{Shukla et~al.(2016)Shukla, Fauve,, and
  Brachet}]{shukla2016statistical}
Shukla, V., S.~Fauve, and M.~Brachet, 2016: Statistical theory of reversals in
  two-dimensional confined turbulent flows. \textit{Phys. Rev. E},
  \textbf{94~(6)}, 061\,101, \doi{10.1103/PhysRevE.94.061101}.

\bibitem[{Tantet et~al.(2015)Tantet, van~der Burgt,, and
  Dijkstra}]{tantet2015early}
Tantet, A., F.~R. van~der Burgt, and H.~A. Dijkstra, 2015: An early warning
  indicator for atmospheric blocking events using transfer operators.
  \textit{Chaos: An Interdisciplinary Journal of Nonlinear Science},
  \textbf{25~(3)}, 036\,406.

\bibitem[{Thiede et~al.(2019)Thiede, Giannakis, Dinner,, and
  Weare}]{thiede2019galerkin}
Thiede, E.~H., D.~Giannakis, A.~R. Dinner, and J.~Weare, 2019: Galerkin
  approximation of dynamical quantities using trajectory data. \textit{J. Chem.
  Phys.}, \textbf{150~(24)}, 244\,111,
  \doi{10.1063/1.5063730@jcp.2019.MMMK.issue-1}.

\bibitem[{Thompson(1957)}]{Thompson1957}
Thompson, P.~D., 1957: Uncertainty of initial state as a factor in the
  predictability of large scale atmospheric flow patterns. \textit{Tellus},
  \textbf{9~(3)}, 275--295, \doi{10.1111/j.2153-3490.1957.tb01885.x}.

\bibitem[{Timmermann and Jin(2002)Timmermann, and
  Jin}]{timmermann2002nonlinear}
Timmermann, A., and F.-F. Jin, 2002: {A nonlinear mechanism for decadal El
  Ni{\~n}o amplitude changes}. \textit{Geophys. Res. Lett.}, \textbf{29~(1)},
  3--1, \doi{10.1029/2001GL013369}.

\bibitem[{Timmermann et~al.(2003)Timmermann, Jin,, and
  Abshagen}]{timmermann2003nonlinear}
Timmermann, A., F.-F. Jin, and J.~Abshagen, 2003: {A nonlinear theory for El
  Ni{\~n}o bursting}. \textit{J. Atmos. Sci.}, \textbf{60~(1)}, 152--165,
  \doi{10.1175/1520-0469(2003)060<0152:ANTFEN>2.0.CO;2}.

\bibitem[{Vanden-Eijnden(2006)}]{vanden2006transition}
Vanden-Eijnden, E., 2006: Transition path theory. \textit{Computer Simulations
  in Condensed Matter Systems: From Materials to Chemical Biology Volume 1},
  Springer, 453--493.

\end{thebibliography}

\clearpage

\end{document}